%% file: main.tex
\documentclass[12pt]{article}
\usepackage{setspace}
\usepackage{amssymb,graphicx,color}
\usepackage{amsfonts}
\usepackage{algorithm}
\usepackage{algpseudocode}
\usepackage{amsmath}
\usepackage{latexsym}
\usepackage{array}
\usepackage{geometry}
\usepackage{caption}
\usepackage{subcaption}
\usepackage[hidelinks]{hyperref}
\usepackage{adjustbox}
\usepackage{helvet}

\usepackage{afterpage}
\usepackage{authblk}
\usepackage[figuresleft]{rotating}

\usepackage{booktabs}
\usepackage{longtable}
\usepackage{pdflscape}

\begin{document}

\title{APHABAMAS: An analytical phantom-based scheme for assessing the accuracy of high-resolution 3D MRI motion-artifact simulations}

\author[1,2]{Tianqi Wu}
\author[1,3]{Hui Zhang} 

\affil[1]{Hawkes Institute, University College London, London, United Kingdom}
\affil[2]{Department of Medical Physics and Biomedical Engineering, University College London, London, United Kingdom}
\affil[3]{Department of Computer Science, University College London, London, United Kingdom}

\date{}
\maketitle
\noindent\textbf{Correspondence:} Tianqi Wu (tianqi.wu.18@ucl.ac.uk)

\vspace{0.5em}

\vspace{0.5em}
\noindent\textbf{Abstract word count:} 248

\vspace{0.5em}
\noindent\textbf{Word count (body text only):} 4992

\clearpage
\section*{Abstract}
\noindent \textbf{Purpose:}
Motion compromises the utility of high-resolution 3D MRI, an established tool in quantitative neuroimaging research. Deep learning-based methods have shown promise for mitigating motion-induced artifacts, but their development typically requires simulated motion-corrupted data. Several open-source tools exist for this task. They share a common formulation but implement this formulation with different algorithms. It is unclear which algorithm implements the formulation more accurately. This study aims to develop a scheme to answer this question.

\noindent \textbf{Methods:}
The essential ingredient of the desired scheme is a ground-truth reference that does not suffer from sampling-induced error. To meet this requirement, the proposed scheme, APHABAMAS, leverages a digital phantom whose image- and Fourier-domain representations can be expressed analytically under arbitrary rigid-body transformations. APHABAMAS is used to quantify sampling-induced errors of three existing algorithms under synthetic and tracking-data-derived motion trajectories at different motion severity levels.

\noindent \textbf{Results:}
All three algorithms produce distinct but visually plausible artifacts, demonstrating the need for a ground-truth reference. APHABAMAS shows that the algorithm utilizing the uniform-to-non-uniform (Type-2) Non-Uniform Fast Fourier Transform (NUFFT) provides the most consistent agreement with the ground-truth reference across motion types and severity levels. For the non-uniform-to-uniform (Type-1) NUFFT-based algorithm, agreement with the ground-truth reference deteriorates substantially with increasing motion severity.

\noindent \textbf{Conclusion:}
APHABAMAS provides a rigorous tool for assessing how accurately algorithms implement the common formulation for simulating motion artifacts in high-resolution 3D MRI. It enables accuracy-based ranking of the three assessed algorithms, thereby facilitating informed selection of the most suitable one for synthesizing motion-corrupted data.

\clearpage
\section{Introduction}
Modern 3D MRI sequences, particularly MPRAGE~\cite{mugler1990three} and SPACE~\cite{mugler2000optimized}, enable high-resolution 3D brain imaging with various contrasts within a feasible time frame. They generate essential data, like T1- and T2-weighted scans, allowing detailed examination of brain anatomy and pathology. Consequently, these techniques have become routine choices for large neuroimaging studies, including the Alzheimer's Disease Neuroimaging Initiative~\cite{jack2008alzheimer, gunter2017adni}, the WU-Minn Human Connectome Project~\cite{van2013wu}, and the UK Biobank~\cite{littlejohns2020uk}.

Despite these advantages, however, high-resolution 3D MRI are prone to motion artifacts~\cite{zaitsev2015motion}, as their acquisition times often exceed the period during which children, elderly individuals, and patients with neurological disorders can remain still. Such artifacts degrade image quality and bias downstream analyses, leading to inaccurate biomarker estimates, e.g., gray matter volume and cortical thickness~\cite{reuter2015head, alexander2016subtle}. Therefore, identifying and either discarding or correcting these low-quality scans is essential.

Recent advances in deep-learning (DL) have driven a growing interest in DL-based approaches for mitigating motion artifacts in 3D MRI~\cite{lee2020deep, spieker2023deep}. Applications include motion-robust brain segmentation through artifact-augmented training~\cite{shaw2020k}, motion-artifact detection in T1-weighted 3D MRI~\cite{fantini2021automatic}, and retrospective motion-artifact correction~\cite{johnson2019conditional, duffy2021retrospective, al2022stacked}. Recent methods incorporate DL into joint image and motion estimation for 3D MRI~\cite{nghiem2026network} and physics-driven motion correction for 3D chemical exchange saturation transfer MRI~\cite{singh2025learning}.

These promising studies highlight a central challenge: curating suitable datasets required to develop these methods. For motion-artifact detection, labels describing the presence or severity of the artifacts require labor-intensive manual rating. For motion correction, many approaches~\cite{johnson2019conditional, duffy2021retrospective} require paired motion-free and motion-corrupted scans, which are rarely available. Even when such pairs exist~\cite{narai2022movement}, the corresponding motion-tracking data are typically missing. Approaches that do not require explicit labels still face limited availability of motion-corrupted data, as most public datasets undergo quality control before release. Consequently, simulated motion-corrupted data are widely used for method development and evaluation, including in most of the aforementioned DL-based motion-mitigation studies, providing known motion trajectories and paired motion-free and motion-corrupted scans~\cite{lee2020deep, spieker2023deep}.

Traditionally, many studies implemented custom motion-artifact simulators but rarely released them publicly. This practice limits reproducibility and has necessitated redundant efforts. In recent years, the community has increasingly embraced open-source development, resulting in several open-source motion-artifact simulators for 3D MRI. For example, the TorchIO~\cite{perez2021torchio} Python library offers a simulator based on the simulation algorithm by Shaw et al.~\cite{shaw2020k}. Duffy et al.~\cite{duffy2021retrospective} also released their simulator for generating data for developing motion-correction models. Reguig et al.~\cite{reguig2022global} extended TorchIO with implementations of additional algorithms. More recently, Kumar et al.~\cite{kumar2025pysimpace} also introduced an open-source toolkit for generating motion-corrupted data, named PySimPace. 

Recent studies demonstrate the value of open-source motion simulators. Loizillon et al.~\cite{loizillon2024automatic} developed a transfer learning method for detecting motion artifacts in clinical 3D T1-weighted scans, with data simulated using both the original~\cite{perez2021torchio} and the extended~\cite{reguig2022global} TorchIO. Similarly, Reddy et al.~\cite{reddy2024gan} used TorchIO to simulate data for training a motion correction model. 

With multiple frameworks and implementations available, prospective users face an obvious question: which framework or implementation to adopt. Beyond usability and documentation quality, simulation accuracy is a key factor to consider. Specifically, existing simulators use algorithms that follow a common formulation in which motion-corrupted k-space signals are generated from a representation of the moving object and subsequently used to reconstruct the scan (see Sec.~\ref{sec:theory:problem_formulation} for details). These algorithms use a motion-free scan of the underlying object as input and construct moving object representations from it. Because such a scan is reconstructed from finitely sampled k-space signals, it provides a discrete, band-limited approximation of the object’s underlying spin density. The discrepancy between the two constitutes sampling error. Crucially, the algorithms differ in how they generate motion-corrupted k-space signals from the motion-free scan. We refer to the resulting simulation error arising from the use of this sampled approximation as sampling-induced error, which may vary across algorithms. To the best of our knowledge, no assessment scheme currently exists for determining which algorithm implements the shared formulation more accurately. This study aims to develop such a scheme.

The key challenge in establishing such a scheme is producing a ground-truth simulation free from sampling-induced error. To address this, we propose representing the underlying object using an analytical phantom---a digital phantom whose image- and Fourier-domain representations have closed-form expressions. This allows us to establish an analytical phantom-based scheme for assessing the accuracy of high-resolution 3D MRI motion-artifact simulations, which we refer to as APHABAMAS. With APHABAMAS, we assess and compare the accuracy of three existing simulation algorithms.

The rest of the paper is organized as follows: Sec.~\ref{sec:theory} elaborates on why existing algorithms result in inevitable errors and presents APHABAMAS along with its mathematical foundations; Sec.~\ref{sec:methods} details how we demonstrate APHABAMAS's necessity and utility; Sec.~\ref{sec:results} presents the results of the necessity demonstration and algorithm assessment; and Sec.~\ref{sec:discussion} summarizes key findings and discusses future work.

\clearpage
\section{Theory}
\label{sec:theory}
This section has two aims: (1) to explain why and how existing simulation algorithms result in errors, and (2) to present APHABAMAS and its theoretical basis for generating the ground-truth simulation.

To this end, Sec.~\ref{sec:theory:problem_formulation} presents the mathematical formulation for the motion‑artifact simulation problem that is adopted by the algorithms we aim to assess. With this formulation, Sec.~\ref{sec:theory:sampling_error} identifies the source of error in existing algorithms. This uncovers a necessary condition for the ground-truth simulation, facilitating the establishment of APHABAMAS presented in Sec.~\ref{sec:theory:scheme}. Following that, to implement APHABAMAS, Sec.~\ref{sec:theory:gt_computation} details how to compute the ground‑truth simulation practically. This also enables the subsequent analysis in Sec.~\ref{sec:theory:existing_algorithm} of how existing algorithms deviate from the ground-truth differently.

\subsection{Formulation of motion-artifact simulation of interest}
\label{sec:theory:problem_formulation}
\begin{figure}
    \centering
    \includegraphics[width=\linewidth]{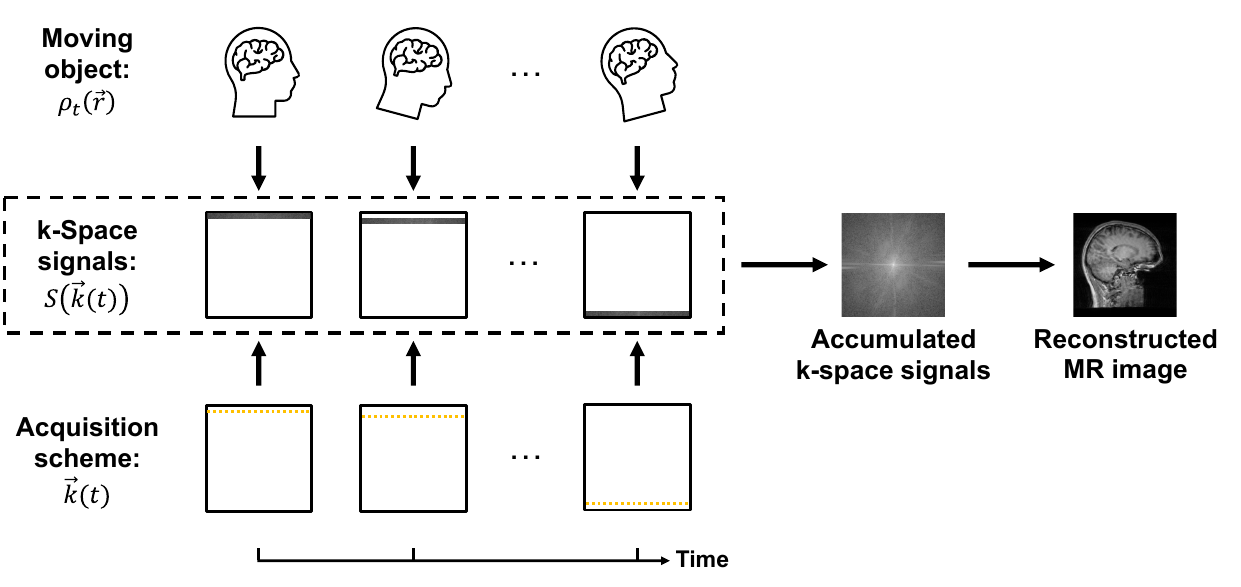}
    \caption{Conceptual overview of the MR acquisition process in the presence of object motion. The interaction between the moving object and the prescribed MR acquisition scheme dictates the acquired k-space signals (the signals are displayed in their log-scaled magnitude). These signals, together with their sampling locations defined by the acquisition scheme, are subsequently used to reconstruct the final motion-corrupted scan.}
    \label{fig:overall_formulation}
\end{figure}

Motion-artifact simulation, in general, mimics MR acquisitions where the imaged object moves during scanning. As illustrated in Fig.~\ref{fig:overall_formulation}, these acquisitions can be described as follows: given a moving object and an acquisition scheme, the scanner acquires MR signals over time, and these signals are subsequently used to reconstruct the motion-corrupted scan.

In line with the mathematical formulation adopted by the algorithms we aim to assess, the moving object is defined by a time‑varying (effective) spin‑density function, $\rho_t(\vec{r})$, which is assumed to account for contrast‑determining factors like proton density and relaxation times. The acquisition scheme is defined by an intended k-space trajectory, $\vec{k}(t)$, and a set of sampling time points, which together specify the signals' k-space sampling locations.

In practice, $\rho_t(\vec{r})$ is parameterized by an initial spin-density function $\rho_0(\vec{r})$ and a time-dependent spatial transformation $A_t(\vec{r})$ that maps each location at time $0$ to its position at time $t$. The transformation is commonly referred to as a motion trajectory. Under the usual assumption that the object moves rigidly, as derived in Supporting Information Sec.~S1.1, $\rho_t(\vec{r})$ can be expressed via $\rho_0(\vec{r})$ and $A_t(\vec{r})$ as
\begin{align}
\rho_t(\vec{r}) = \rho_0\big(A_t^{-1}(\vec{r})\big)\, , \label{eq:rho_t}
\end{align}
where $A_t^{-1}$ denotes the inverse of $A_t$. Because a 3D rigid transformation is uniquely determined by six independent parameters (three for translation and three for rotation), the motion trajectory is typically specified as the temporal evolution of these parameters. In this study, rotation parameters $\theta_{x,t}$, $\theta_{y,t}$, and $\theta_{z,t}$, corresponding to rotations about the left-right, anterior-posterior, and inferior-superior axes, respectively, are combined using $\mathbf{R}_t=\mathbf{R}_z(\theta_{z,t})\mathbf{R}_x(\theta_{x,t})\mathbf{R}_y(\theta_{y,t})$, with constructions of the rotation matrices in Sec.~S1.2.

In 3D MRI, a typical acquisition utilizes multiple gradients to sample k-space on a discrete 3D Cartesian grid, with each readout filling a 1D grid. The complete $\vec{k}(t)$ associated with each readout is achieved by first applying phase-encoding and dephasing frequency-encoding gradients to reach the target position, immediately followed by a frequency-encoding gradient of opposite polarity for sampling. Because these gradients have millisecond-scale durations, it is commonly assumed that object motion during gradient-on intervals has negligible effects. This obviates modeling intra-readout motion.

Following the above definitions and assumptions, the simulated signal can be expressed as:
\begin{align}
\label{eq:signal}
S(\vec{k}(t)) = \int \rho_t(\vec{r}) \, e^{-i 2 \pi \vec{k}(t) \cdot (\vec r-\vec{r}_{iso,t})} \, d\vec{r} {\, ,}
\end{align}
where $\vec{r}_{iso,t}$ represents the isocenter location at time $t$, corresponding to the physical position where the main magnetic field strength remains invariant regardless of gradient activity. In the literature, $\vec{r}_{iso,t}$ is often omitted by adopting a reference frame in which the isocenter is fixed at the origin. We retain it here without committing to a specific choice of reference frame, and the reason will become apparent in Sec.~\ref{sec:theory:gt_computation}. Substituting $\rho_t(\vec{r})$ using Eq.~\eqref{eq:rho_t} yields:
\begin{align}
\label{eq:signal_using_At}
S(\vec{k}(t)) = \int \rho_0(A_t^{-1}(\vec{r})) \, e^{-i 2 \pi \vec{k}(t) \cdot (\vec r-\vec{r}_{iso,t})} d\vec{r} {\, .}
\end{align}

Once the signals required by the acquisition scheme are simulated, the scan can be reconstructed. Because the sampling locations form a Cartesian grid, applying an inverse Fast Fourier Transform (IFFT) and taking its magnitude yields the final simulation.

In practice, however, when implementing the above formulation, all existing algorithms introduce inevitable errors, as we will explain next.

\subsection{Source of error in existing algorithms}
\label{sec:theory:sampling_error}
Under the above formulation, the fundamental source of error in current simulation algorithms stems from inaccuracies in the initial spin-density of the object, $\rho_0(\vec{r})$. This is because existing algorithms utilize a motion-free MR scan as a proxy for $\rho_0(\vec{r})$, given that direct access to the brain's true $\rho_0(\vec{r})$ is unavailable. Such a scan, being discretely sampled and reconstructed from finitely sampled k-space signals, is inherently a discrete and band-limited approximation of the true $\rho_0(\vec{r})$. Consequently, the motion-corrupted signals derived from it via Eq.~\eqref{eq:signal_using_At} are themselves approximate. This leads to error, which we have defined as sampling-induced error. Furthermore, because each algorithm uses the motion-free scan differently, their resulting errors vary (see Sec.~\ref{sec:theory:existing_algorithm}), underscoring the need for developing an assessment scheme.

The presence of sampling-induced error in current algorithms reveals a key requirement for generating the ground-truth simulation needed for an assessment scheme: the availability of the true underlying $\rho_0(\vec{r})$. This requirement guides the scheme developed below.

\subsection{Proposed ground-truth-enabled scheme: APHABAMAS}
\label{sec:theory:scheme}
\begin{figure}
    \centering
    \includegraphics[width=\linewidth]{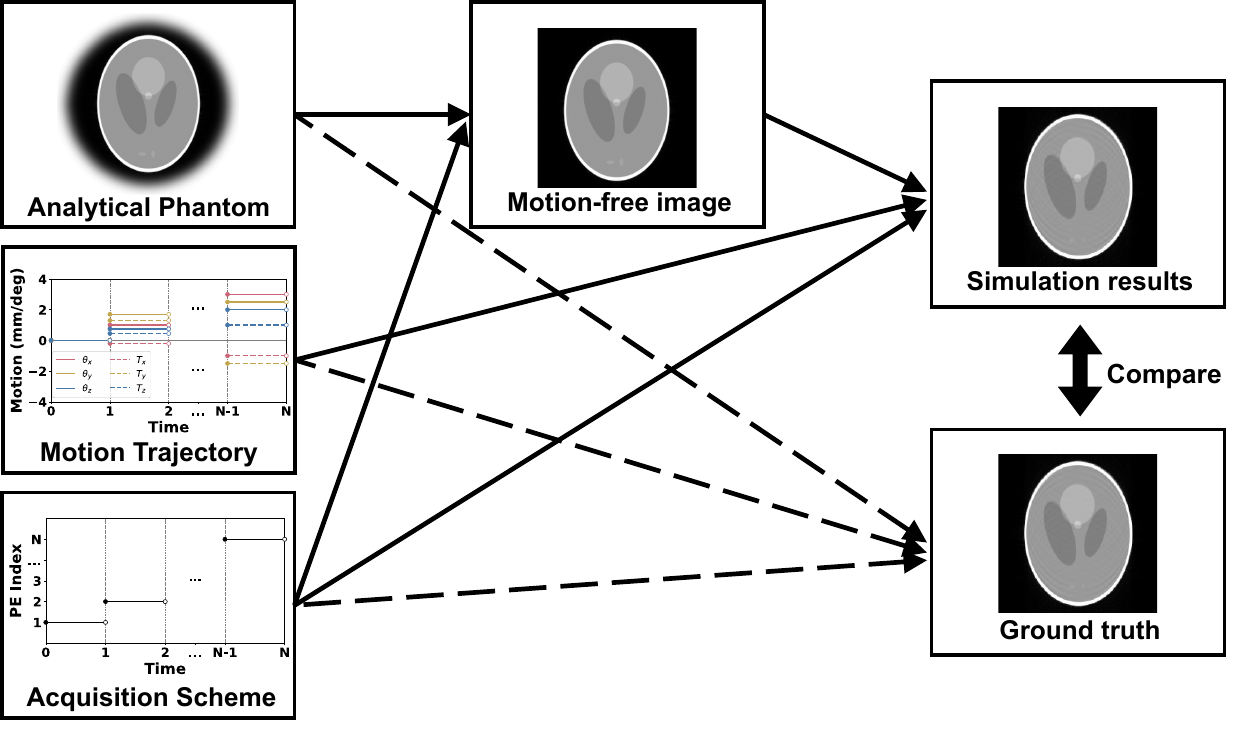}
    \caption{Overview of APHABAMAS. The analytical ground truth (dashed path) and the algorithm-simulated output (solid path) are generated using identical motion and acquisition parameters. The algorithm-simulated output is then assessed by comparison with the ground truth.}
    \label{fig:assessment_scheme}
\end{figure}

The proposed ground-truth-enabled assessment scheme, APHABAMAS (Fig.~\ref{fig:assessment_scheme}), evaluates an algorithm by comparing its output against a ground-truth simulation. Establishing the ground truth first requires the availability of $\rho_0(\vec{r})$, as highlighted in Sec.~\ref{sec:theory:sampling_error}. However, this alone is insufficient; exactly computing $S(\vec{k}(t))$ via Eq.~\eqref{eq:signal_using_At} requires the Fourier transform of $\rho_0(A_t^{-1}(\vec{r}))$  to have a closed-form expression. Because $A_t^{-1}(\vec{r})$ is a rigid transformation, this should naturally be satisfied if the Fourier transform of $\rho_0(\vec{r})$ has a closed-form expression. We identify that representing the underlying object with an analytical phantom, whose image- and Fourier-domain representations possess closed-form expressions, satisfies both requirements simultaneously. Such a phantom enables computing the ground-truth simulation (following Sec.~\ref{sec:theory:gt_computation}). Concurrently, such a phantom allows for the reconstruction of its motion-free scan, serving as the proxy of $\rho_0(\vec{r})$ required by existing algorithms.

The literature offers two suitable phantoms: (i) the 3D Shepp–Logan phantom~\cite{shepp1974fourier, kak2001principles}, comprising overlapping ellipsoids of varying sizes and shapes, which Koay et al.~\cite{koay2007three} proved to meet our requirements; and (ii) a polyhedral brain phantom developed by Ngo et al.~\cite{ngo2016realistic} to offer greater anatomical realism while meeting the same requirements we desire. Because the latter's computational cost is prohibitive, approximately 0.4s to simulate a k-space sample, and extreme anatomical realism is unnecessary for our purpose, this study uses the former choice.

\subsection{Practical computation of ground-truth simulation}
\label{sec:theory:gt_computation}
The computation of the ground-truth simulation involves calculating $S(\vec{k}(t))$ via Eq.~\eqref{eq:signal_using_At}. This typically requires choosing a specific reference frame so that $\vec{k}(t)$, $\vec{r}$, and $A_t(\vec{r})$ can be formulated in terms of objects of linear algebra. Although all frames are equivalent, we consider two specific choices because they are used in some of the existing simulation algorithms. Presenting both allows us to dissect, in Sec.~\ref{sec:theory:existing_algorithm}, the differing consequences of these algorithms that require the proxy of $\rho_0(\vec{r})$ as input. 

The most intuitive approach is to define $\vec{k}(t)$, $\vec{r}$, and $A_t(\vec{r})$ within a stationary reference frame fixed to the scanner. This corresponds to viewing motion from the scanner's perspective. Conversely, the other approach adopts the object's local frame of reference. This is essentially viewing the motion from the object's perspective, where the object remains stationary while the scanner moves.

\subsubsection{Describing motion from scanner's perspective}
\label{sec:theory:gt_computation:scanner_frame}
To describe motion from the scanner's perspective, the scanner's frame of reference is adopted. By scanner frame, we mean a specific right-handed Cartesian coordinate system $(x, y, z)$ centered at the magnet's isocenter. We denote spatial coordinates of a point in the scanner frame as $\mathbf{r}^s=[x^s, y^s, z^s]^\top$, with the coordinates in its corresponding k-space denoted as $\mathbf{k}^s(t)=[k_x^s(t), k_y^s(t), k_z^s(t)]^\top$.

Because the isocenter of the scanner frame aligns with the origin ($\mathbf{r}^s_{iso,t}=[0,0,0]^\top$), the MR signal at time $t$ (Eq.~\eqref{eq:signal_using_At}) becomes:
\begin{align}
S(\mathbf{k}^s(t)) =\int \rho_0(A_t^{-1}(\mathbf{r}^s)) e^{-i 2\pi\, \mathbf{k}^s(t) \cdot (\mathbf{r}^s-\mathbf{r}^s_{iso,t})}\, d\mathbf{r}^s
=\int \rho_0(A_t^{-1}(\mathbf{r}^s)) e^{-i 2\pi\, \mathbf{k}^s(t) \cdot \mathbf{r}^s}\, d\mathbf{r}^s {\, .}
\label{eq:signal_scanner_frame}
\end{align}
Here, $A_t$ is defined in the scanner frame as:
\begin{align}
A_t(\mathbf{r}^s) = \mathbf{R}_t \mathbf{r}^s + \mathbf{T}_t {\, ,}
\label{eq:rigid}
\end{align}
where $\mathbf{R}_t \in \mathbb{R}^{3 \times 3}$ is the rotation matrix and $\mathbf{T}_t \in \mathbb{R}^{3}$ is the translation vector. Substituting the inverse transformation $A_t^{-1}(\mathbf{r}^s) = \mathbf{R}_t^{\top}(\mathbf{r}^s - \mathbf{T}_t)$ into Eq.~\eqref{eq:signal_scanner_frame} yields:
\begin{align}
S(\mathbf{k}^s(t)) = \int \rho_0(\mathbf{R}_{t}^{\top}(\mathbf{r}^s - \mathbf{T}_{t}))
  e^{-i 2\pi\, \mathbf{k}^s(t) \cdot \mathbf{r}^s}\, d\mathbf{r}^s {\, .}
\label{eq:FT_of_rho_scanner_frame}
\end{align}

\begin{figure}
    \centering
        \includegraphics[width=0.85\linewidth]{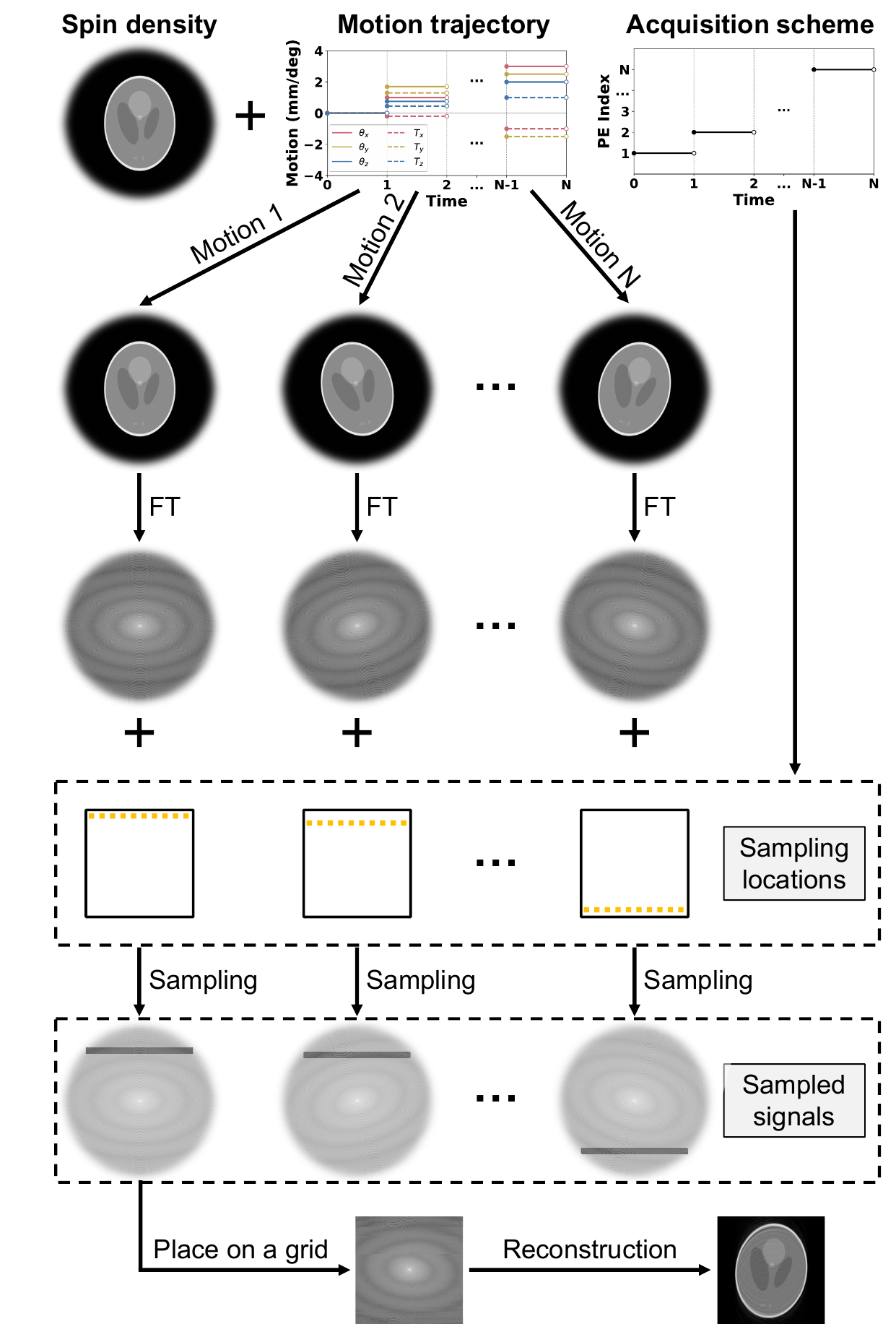}
    \caption{Visualization of ground-truth simulation using the scanner-frame formulation. The process is illustrated in 2D for clarity. For motion trajectory and acquisition scheme, the y-axis tick labels $1,\ldots,N$ mark the end times of successive readouts, where $N$ is the total number of phase-encoding (PE) steps. Circles with blurry edges represent the underlying object in real space or k-space, while black square boxes indicate the k-space fields of view when showing sampling locations. The k-space representations and their sampled versions are displayed in log-scaled magnitude.}
    \label{fig:scanner_frame}
\end{figure}

Simulation using Eq.~\eqref{eq:FT_of_rho_scanner_frame} is illustrated in Fig.~\ref{fig:scanner_frame}. Specifically, the transformed spin density $\rho_0(\mathbf{R}_{t}^{\top}(\mathbf{r}^s - \mathbf{T}_{t}))$ corresponding to each readout is computed. The Fourier transform of each $\rho_0(\mathbf{R}_{t}^{\top}(\mathbf{r}^s - \mathbf{T}_{t}))$ is then calculated. Required signals are then evaluated at the k-space locations specified by the acquisition trajectory for each readout, using the Fourier transform of the corresponding $\rho_0(\mathbf{R}_{t}^{\top}(\mathbf{r}^s - \mathbf{T}_{t}))$. Finally, these signals are gridded based on sampling locations and used for reconstruction.

\subsubsection{Describing motion from object's perspective}
\label{sec:theory:gt_computation:object_frame}
Alternatively, to describe motion from the object's perspective, the object frame is adopted. By the object frame, we refer to a moving coordinate system in which the object remains stationary throughout the acquisition. It coincides with the scanner frame at $t=0$ but subsequently translates and rotates alongside the object. We denote spatial coordinates in the object frame as $\mathbf{r}^o=[x^o, y^o, z^o]^\top$, with the coordinates in its corresponding k-space being $\mathbf{k}^o(t)=[k_x^o(t), k_y^o(t), k_z^o(t)]^\top$.

Because the object remains fixed in this frame ($\rho_t(\mathbf{r}^o) = \rho_0(\mathbf{r}^o)$ for all $t$), the signal (Eq.~\eqref{eq:signal}) expressed in object-frame coordinates becomes:
\begin{align}
S(\mathbf{k}^o(t)) &= \int \rho_t(\mathbf{r}^o) e^{-i 2\pi\, \mathbf{k}^o(t) \cdot (\mathbf{r}^o - \mathbf{r}^o_{iso,t})}\, d\mathbf{r}^o =\int \rho_0(\mathbf{r}^o) e^{-i 2\pi\, \mathbf{k}^o(t) \cdot (\mathbf{r}^o - \mathbf{r}^o_{iso,t})}\, d\mathbf{r}^o {\, .}
\label{eq:signal_object_frame}
\end{align}
Calculating signals via Eq.~\eqref{eq:signal_object_frame} requires two factors to be resolved. First, because k-space sampling locations are conventionally specified in scanner frame coordinates ($\mathbf{k}^s(t)$), the relation between $\mathbf{k}^s(t)$ and $\mathbf{k}^o(t)$ needs to be determined. Second, in the object frame, the isocenter no longer always coincides with the origin; hence the expression for $\mathbf{r}^o_{iso,t}$ is required.

The relationship between $\mathbf{k}^s(t)$ and $\mathbf{k}^o(t)$ can be deduced from two principles of MR physics: (1) the k-space axes are dictated by applied gradient directions; and (2) rotating the object by $\mathbf{R}_t$ is equivalent to rotating the gradients in the opposite direction by $\mathbf{R}^\top_t$. Consequently, the object-frame k-space axes are rotated by $\mathbf{R}^\top_t$ relative to the scanner frame, yielding:
\begin{align}
\label{eq:kspacce_frame_change}
\mathbf{k}^o(t) = \mathbf{R}^\top_t\mathbf{k}^s(t)
\end{align}

To express the isocenter $\mathbf{r}^o_{iso,t}$, we must determine the relationship between $\mathbf{r}^o$ and its corresponding spatial coordinates within the scanner frame at time $t$, which we denote as $\mathbf{r}^s_t$. For any given point, its object-frame coordinates $\mathbf{r}^o$ remain fixed. Because the object and scanner frames align at time $t=0$, we merely need to determine the relation between a point's coordinates in the scanner frame at time $t=0$ and its coordinates (also in the scanner frame) at time $t$. The transformation $A_t$ defined previously in Sec.~\ref{sec:theory:gt_computation:scanner_frame} does exactly that; thus:
\begin{align}
\label{eq:frame_change}
\mathbf{r}^s_t = A_t(\mathbf{r}^s_0)= A_t(\mathbf{r}^o) = \mathbf{R}_t\mathbf{r}^o + \mathbf{T}_t {\, .}
\end{align}
As a result, $\mathbf{r}^o_{iso,t}$ can be expressed using $\mathbf{r}^s_{iso,t}$:
\begin{align}
\label{eq:isocenter}
\mathbf{r}^o_{iso,t} = \mathbf{R}^\top_t(\mathbf{r}^s_{iso,t} - \mathbf{T}_t) {\, .}
\end{align}
Given that the chosen scanner frame in Sec.~\ref{sec:theory:gt_computation:scanner_frame} sets the isocenter at the origin ($\mathbf{r}^s_{iso,t}=[0,0,0]^\top$) for the scanner frame, this simplifies to:
\begin{align}
\label{eq:isocenter_2}
\mathbf{r}^o_{iso,t} = -\mathbf{R}^\top_t\mathbf{T}_t {\, .}
\end{align}

With Eq.~\eqref{eq:kspacce_frame_change} and Eq.~\eqref{eq:isocenter_2}, the desired signal (Eq.~\eqref{eq:signal_object_frame}) can be rewritten in terms of $\mathbf{k}^s(t)$ as:
\begin{align}
\label{eq:signal_object_frame_final}
S(\mathbf{k}^s(t)) &= \int \rho_0(\mathbf{r}^o) e^{-i 2\pi (\mathbf{R}^\top_t \mathbf{k}^s(t)) \cdot (\mathbf{r}^o + \mathbf{R}^\top_t\mathbf{T}_t)} d\mathbf{r}^o \nonumber \\
&= e^{-i 2\pi (\mathbf{R}^\top_t \mathbf{k}^s(t)) \cdot (\mathbf{R}^\top_t\mathbf{T}_t)} \int \rho_0(\mathbf{r}^o) e^{-i 2\pi (\mathbf{R}_{t}^\top \mathbf{k}^s(t)) \cdot \mathbf{r}^o} d\mathbf{r}^o \nonumber \\
&= \underbrace{e^{-i 2\pi \mathbf{k}^s(t) \cdot \mathbf{T}_{t}}}_{\text{phase ramp}} S_0(\mathbf{R}_{t}^\top \mathbf{k}^s(t)) {\, ,}
\end{align}
where $S_0(\mathbf{k}^s)$ is the Fourier transform of $\rho_0(\mathbf{r}^s)$. This demonstrates that in the object frame, motion manifests as a rotation of the effective sampling location together with a linear phase ramp due to translation.

\begin{figure}
    \centering
        \includegraphics[width=0.85\linewidth]{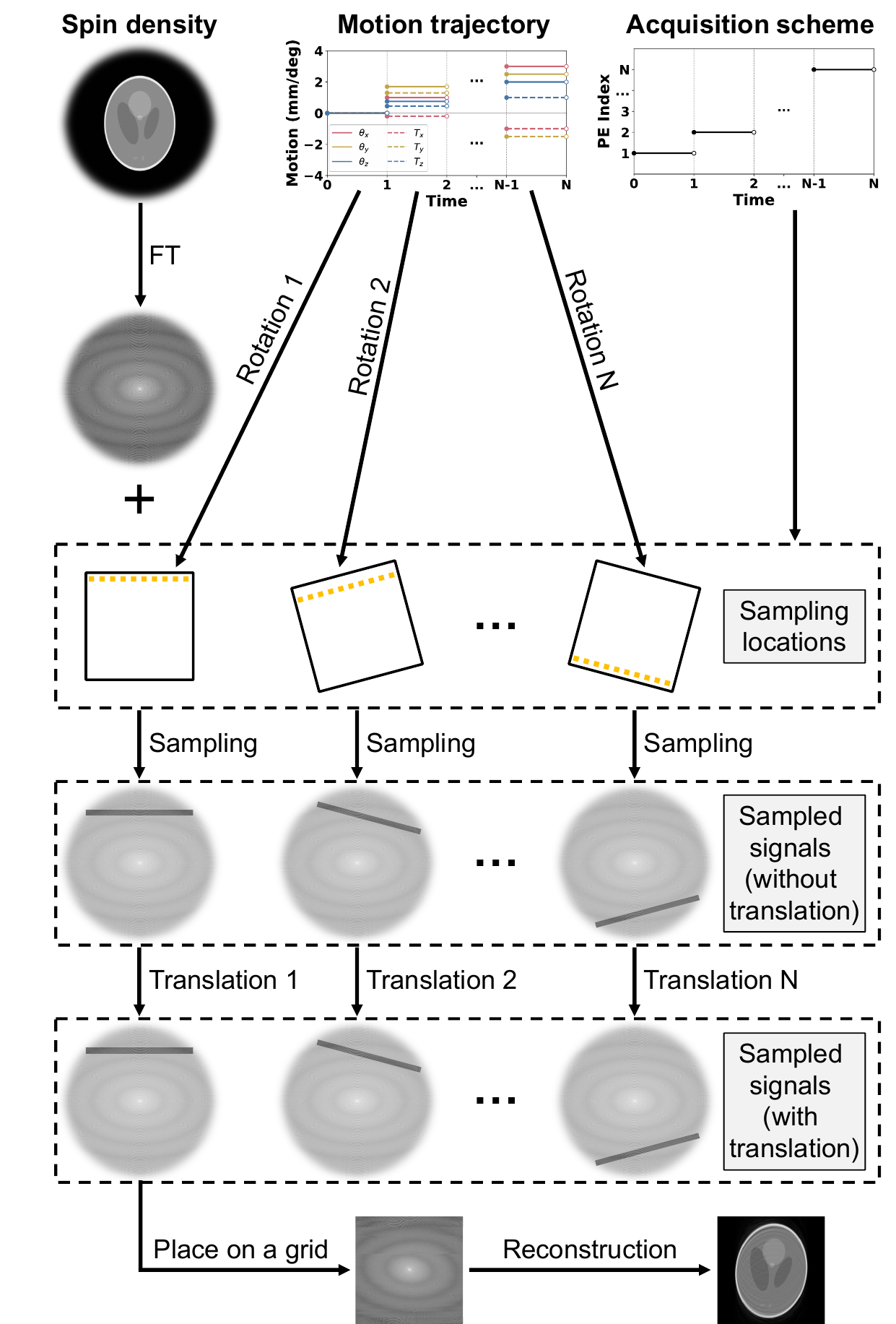}
    \caption{Visualization of ground-truth simulation using the object-frame formulation. The process is illustrated in 2D for clarity, following the plotting conventions in Fig.~\ref{fig:scanner_frame}. The sampling locations appear rotated at different time points because they are expressed in the object frame, rather than because of changes to the prescribed acquisition trajectory. The k-space representations and their sampled versions are displayed in log-scaled magnitude.}
    \label{fig:object_frame}
\end{figure}

The simulation process using this approach is illustrated in Fig.~\ref{fig:object_frame}. First, the analytical expression for $S_0(\mathbf{k}^s)$ is obtained by applying a Fourier transform to $\rho_0(\mathbf{r}^o)$. Next, using the k-space sampling trajectory and the motion trajectory (specifically the rotation parameters), the effective sampling locations $\mathbf{R}_{t}^\top \mathbf{k}^s(t)$ are calculated. These modified locations and the expression for $S_0(\mathbf{k})$ enable the exact calculation of $S_0(\mathbf{R}_t^\top\mathbf{k}^s(t))$ via Eq.~\eqref{eq:signal_object_frame_final}. A phase ramp is then computed for each readout based on the translational motion parameters and applied to the intermediate sampled signals, yielding the final motion-corrupted signals for image reconstruction.

\subsection{Algorithm-specific deviations from ground truth}
\label{sec:theory:existing_algorithm}
We next analyze how three existing algorithms deviate from the ground truth: the image-based, Type-2 Non-Uniform Fast Fourier Transform (NUFFT)-based, and Type-1 NUFFT-based algorithms. NUFFTs efficiently approximate the corresponding Non-Uniform Discrete Fourier Transforms (NUDFTs). Type-2 maps samples on a uniform grid to outputs at non-uniform locations, whereas Type-1 performs the reverse operation. Their specific uses are described in Secs.~\ref{sec:theory:existing_algorithm:type2} and \ref{sec:theory:existing_algorithm:type1}, respectively. The image-based and Type-2 NUFFT-based algorithms conceptually follow the formulation in Sec.~\ref{sec:theory:gt_computation} but approximate the true $\rho_0(\vec{r})$ with a motion-free scan. We examine how their different uses of this approximate input lead to different deviations from the ground truth. Because the Type-1 NUFFT-based algorithm does not mimic actual MR acquisition, we analyze the additional deviations arising from this theoretical inconsistency.

\subsubsection{Image-based algorithm}
\begin{figure}
    \centering
        \includegraphics[width=0.85\linewidth]{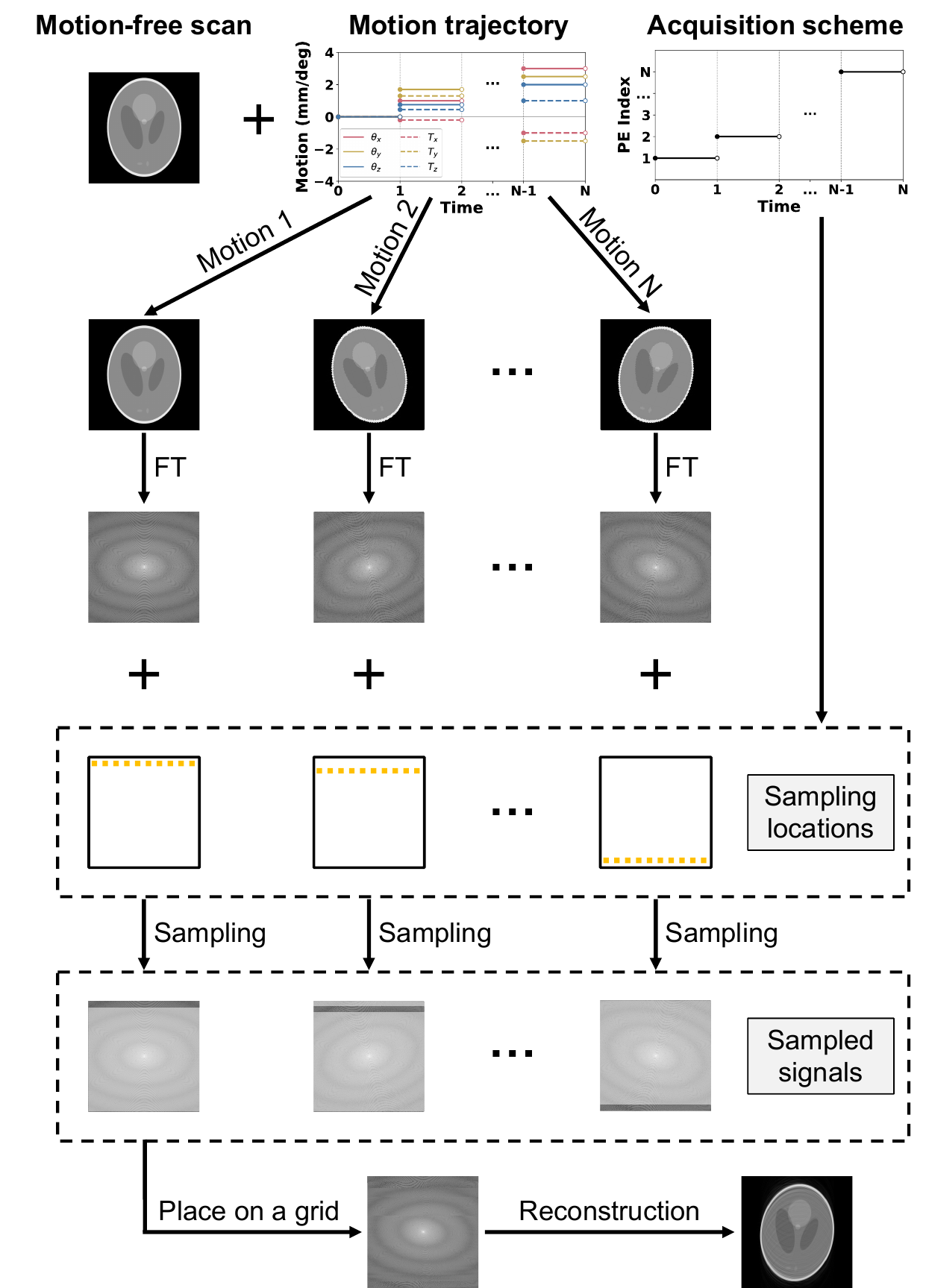}
    \caption{Visualization of the image-based motion-artifact simulation algorithm in 2D. This algorithm represents a practical, discrete implementation of the scanner-frame formulation illustrated in Fig.~\ref{fig:scanner_frame}, using a motion-free scan as a proxy for the underlying spin density. The time indexing for the motion trajectory and acquisition scheme follows Fig.~\ref{fig:scanner_frame}.}
    \label{fig:image_based_algorithm}
\end{figure}

The image-based algorithm conceptually follows the scanner-frame-based formulation (Sec.~\ref{sec:theory:gt_computation:scanner_frame}). Comparing its illustration in Fig.~\ref{fig:image_based_algorithm} against Fig.~\ref{fig:scanner_frame}, substituting the object spin density with a motion-free scan affects computing the transformed spin densities, $\rho_0(\mathbf{R}_{t}^{\top}(\mathbf{r}^s - \mathbf{T}_{t}))$. This introduces errors for three reasons. 

First, the unavailability of true $\rho_0(\vec{r})$ prevents access to different $\rho_0(\mathbf{R}_{t}^{\top}(\mathbf{r}^s - \mathbf{T}_{t}))$, and these, as visualized in Fig.~\ref{fig:image_based_algorithm}, are replaced with discrete, band-limited scans. Second, these scans are constructed from the motion-free scan using linear interpolation, introducing interpolation errors. This is because the motion-free scan only provides values at fixed voxel centers, the transformed coordinates $\mathbf{R}_t^\top(\mathbf{r}^s - \mathbf{T}_t)$ generally land off-grid; estimating values at these locations requires linear interpolation. Finally, because the voxel values of the motion-free scan are themselves band-limited approximations, they inject additional errors that compound the interpolation error. Ultimately, these errors propagate to every sampled $S(\mathbf{k}^{s}(t))$, degrading the resulting simulation.

\subsubsection{Type-2 NUFFT-based algorithm}
\label{sec:theory:existing_algorithm:type2}
\begin{figure}
    \centering
        \includegraphics[width=0.85\linewidth]{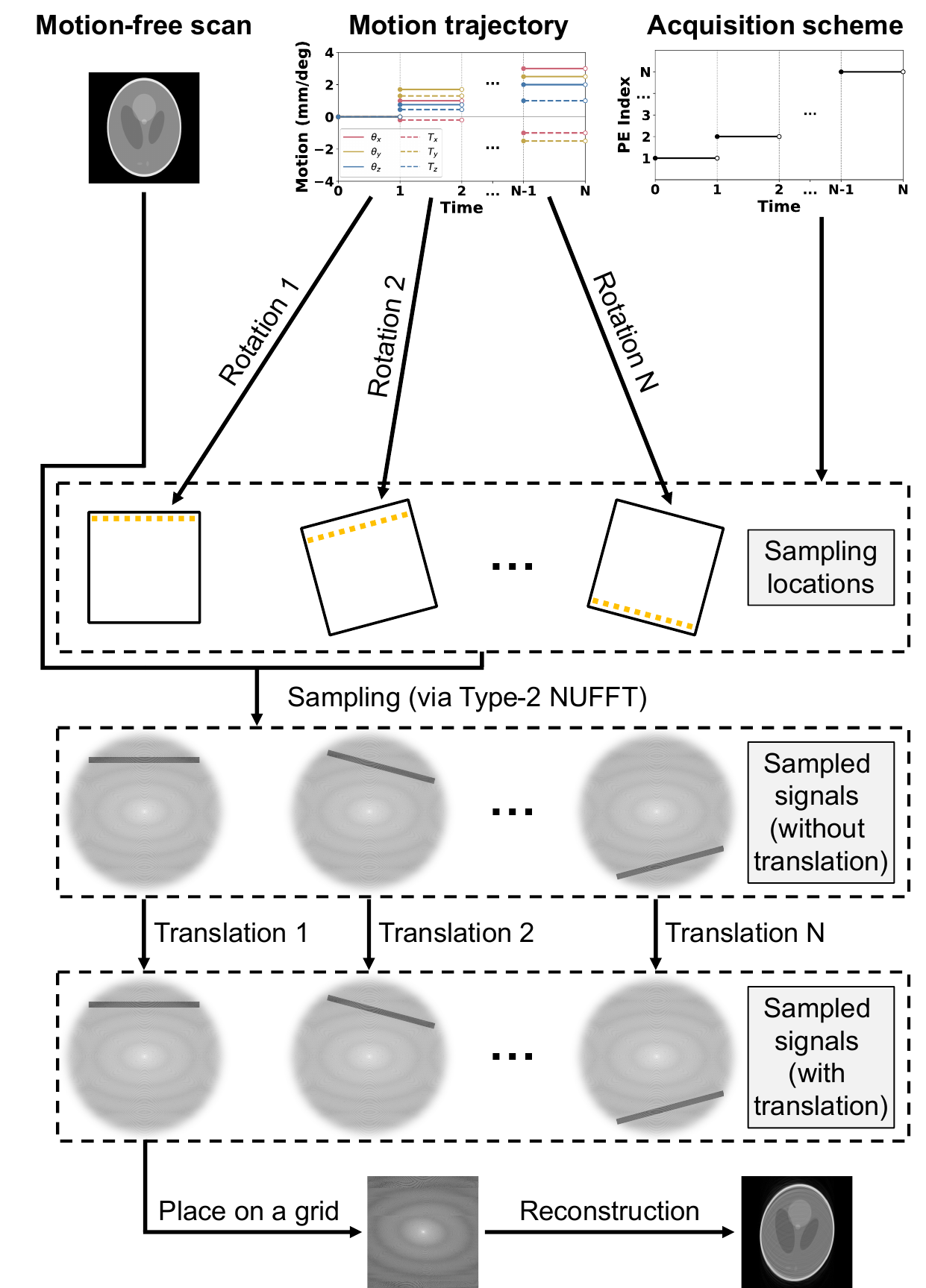}
    \caption{Visualization of the Type-2 NUFFT-based motion-artifact simulation algorithm in 2D. This algorithm represents a practical, discrete implementation of the object-frame formulation illustrated in Fig.~\ref{fig:object_frame}, using a motion-free scan as a proxy for the underlying spin density. The time indexing for the motion trajectory and acquisition scheme follows the same convention as previous figures. The sampling locations appear rotated because they are expressed in the object frame, rather than because of changes to the prescribed acquisition trajectory. The Type-2 NUFFT estimates signals at the required non-uniformly spaced locations. As in Fig.~\ref{fig:object_frame}, the corresponding sampling spaces are illustrated with rounded shapes. However, unlike the ground-truth formulation, these sampling spaces are periodic, as explained in Sec.~\ref{sec:theory:existing_algorithm:type2}.}
    \label{fig:type2_nufft_algorithm}
\end{figure}

The Type-2 NUFFT-based algorithm conceptually follows the object-frame formulation (Sec.~\ref{sec:theory:gt_computation:object_frame}). A comparison of its illustration in Fig.~\ref{fig:type2_nufft_algorithm} against Fig.~\ref{fig:object_frame} shows that substituting the object spin density with a motion-free scan fundamentally alters signal computation. Instead of directly evaluating signals from the Fourier transform of the true object spin density, this algorithm applies a Type-2 NUFFT to the motion-free scan to evaluate the required signals at the motion-modified sampling locations. This introduces simulation errors in three ways.

First, because the motion-free scan is reconstructed from the actual sampled signals, as explained in Sec.~S1.3, calculating signals at non-grid locations from the motion-free scan via Type-2 NUDFT acts as a k-space interpolation using the actual sampled signals, introducing interpolation errors. Second, accelerating NUDFT using NUFFT introduces numerical errors through internal gridding and interpolation steps. Finally, the discrete input image enforces periodicity in the k-space spectrum evaluated by the NUDFT, which means querying a coordinate $\mathbf{k} = (k_x, k_y, k_z)$ beyond the original k-space field-of-view (FOV) causes spectral wraparound (e.g., a $k_x$ exceeding the FOV yields the signal value at $k_x - \text{FOV}_{k_x}$). Because rotational motion causes a small portion of sampling locations to move outside the original k-space FOV, minor additional errors are introduced.

\subsubsection{Type-1 NUFFT-based algorithm}
\label{sec:theory:existing_algorithm:type1}
The Type-1 NUFFT-based algorithm fundamentally differs because it directly manipulates signals derived from the motion-free scan rather than modeling actual acquisition. We focus on explaining its simulation process to understand the error arising from this theoretical inconsistency.

As visualized in Fig.~S1 in the Supporting Information, this algorithm first applies an FFT to the motion-free scan to derive the motion-free signals. To simulate translations, phase ramps are applied, as in the Type-2 NUFFT-based algorithm. Rotations are handled differently: the algorithm assumes that a signal sampled at $\mathbf{k}$ without motion is instead sampled at $\mathbf{R}_t \mathbf{k}$ due to rotation. Reassigning signals to those rotated locations breaks the Cartesian grid structure; therefore, the resulting signals are reconstructed using a Type-1 NUFFT.

\clearpage
\section{Methods}
\label{sec:methods}
This section demonstrates the necessity and utility of APHABAMAS through two experiments. First, we compare algorithms using a real brain scan to reveal that (1) existing algorithms exhibit substantial discrepancies, motivating the need for assessment, and (2) all simulations appear visually realistic, underscoring the need for a ground-truth reference. Second, we demonstrate APHABAMAS's utility by deploying it to evaluate and rank each algorithm. We re-implement the three assessed algorithms for our experiments to have better control and to eliminate implementation-specific confounders. Below, we outline the data and simulation parameters used (Sec.~\ref{sec:methods:data}), followed by the necessity (Sec.~\ref{sec:methods:discrepancies}) and utility (Sec.~\ref{sec:methods:accuracy}) demonstrations of APHABAMAS.

\subsection{Data and simulation parameters}
\label{sec:methods:data}
As described in Sec.~\ref{sec:theory:existing_algorithm}, the assessed algorithms require three inputs: a motion-free scan, a motion trajectory, and an acquisition scheme. Their setups are detailed below.

\subsubsection{Motion-free scan}
\label{sec:methods:data:scan}
The motion-free brain scan used is obtained from the MR-ART dataset~\cite{narai2022movement} ($1~\mathrm{mm}^3$ isotropic voxels, $192 \times 256 \times 256$ matrix size). Its intensities, treated as effective spin-density values, are first capped at the 99th percentile to enhance the visibility of brain structures and simulated motion artifacts, then normalized to $[0, 2]$ to match the range of the analytical phantom~\cite{koay2007three}, enabling comparisons between simulations generated from both. 

The motion-free phantom scan required by APHABAMAS is generated with the same voxel and matrix sizes as the brain scan. These parameters, together with the prescribed acquisition scheme (Sec.~\ref{sec:methods:data:acquisition_scheme}), determine the 3D k-space sampling locations following standard MRI convention (Sec.~S1.5). The analytical Fourier transform of the phantom is then evaluated at these locations to obtain the corresponding k-space signals, which are subsequently reconstructed using IFFT followed by taking their absolute value to obtain the motion-free image.

\subsubsection{Acquisition scheme}
\label{sec:methods:data:acquisition_scheme}
The acquisition scheme used for all experiments is derived from the MPRAGE protocol of MR-ART~\cite{narai2022movement}. Its readout direction aligns with the superior–inferior axis, with each readout filling a 1D grid along this axis. The 3D k-space grid is filled by collecting readouts first along the left–right axis to form a planar grid, and then along the anterior–posterior axis to complete the full 3D grid. Under-sampling for parallel imaging with a multi-coil setup in the original protocol is omitted as it is outside this study's scope.

\subsubsection{Motion trajectories}
\label{sec:methods:data:motion_trajectory}
Both synthetic and tracking-data-derived motion trajectories are used. In both cases, we specify one parameter set per TR instead of the unique per-readout parameters as described in Sec.~\ref{sec:theory:problem_formulation}. This essentially assumes the object remains stationary not only within each readout but across an entire TR (i.e., while filling a planar grid). This common simplification is sufficient for algorithm comparison.

We use synthetic trajectories to examine algorithm performance under four representative motion patterns: slow drift, stepwise motion, spikes, and random jitter. Such motion characteristics have been considered in previous MRI motion studies~\cite{ke2026motion, schauman2026exploration}. For each pattern, we generated 20 trajectories at each of three severity levels (mild, moderate, and severe), with generation procedures and severity definitions detailed in Supporting Information Sec.~S1.4. Representative synthetic trajectories are shown in Fig.~S2.

\begin{figure}[t]
    \centering
        \includegraphics[width=\linewidth]{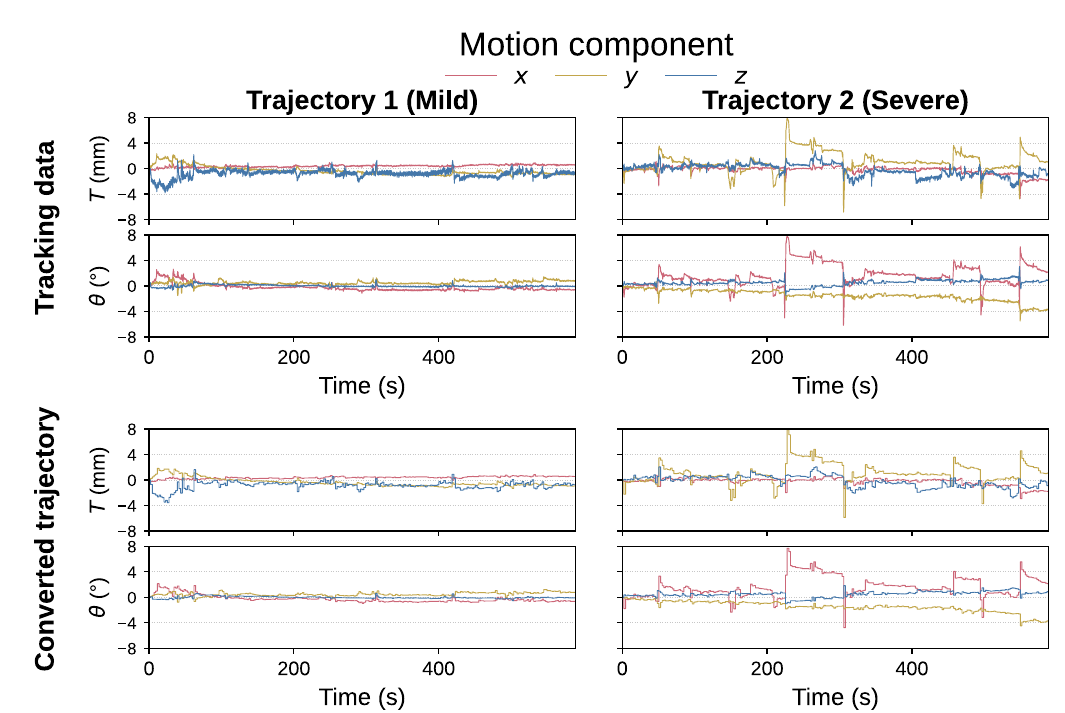}
    \caption{Representative tracking-data-derived motion trajectories used in the experiments. The original tracking measurements and the corresponding trajectories obtained through per-TR temporal discretization are shown for two cases exhibiting relatively mild and severe motion. Rotational parameters follow the convention $\mathbf{R}=\mathbf{R}_z\mathbf{R}_x\mathbf{R}_y$ defined in Sec.~\ref{sec:theory:problem_formulation} and Supporting Information Sec.~S1.2.}
    \label{fig:motion_trajectories}
\end{figure}

Tracking-data-derived trajectories are constructed from a public motion-tracking database~\cite{skare2024brainmotion}. Eight datasets are selected based on having no out-of-FOV motion and sufficient length to cover a 588.8~s scan duration. This duration was calculated using the MR-ART protocol's 2300~ms repetition time (TR) and 256 phase-encoding steps, assuming full k-space sampling because the under-sampling used in the original protocol is not modeled here. The tracking data were converted to motion trajectories using a per-TR temporal discretization. The selected datasets qualitatively span a range of motion severity. Fig.~\ref{fig:motion_trajectories} shows the original tracking data and corresponding motion trajectories for two cases with different motion levels, with the remaining cases shown in Fig.~S3.

\subsection{Demonstrating necessity of APHABAMAS}
\label{sec:methods:discrepancies}
To demonstrate the necessity of APHABAMAS, we evaluate numerical discrepancies and visual realism of simulations across different algorithms. Artifact-corrupted scans are simulated using the motion-free brain scan and the trajectories from Sec.~\ref{sec:methods:data:motion_trajectory}. Pairwise inter-algorithm discrepancies are quantified via two volume-wise metrics: the Structural Similarity Index Measure (SSIM) for quantifying the similarity of ghosting and blurring patterns, and the Root Mean Square Deviation (RMSD) for measuring numerical intensity differences. Both are computed over all voxels to include foreground and background artifacts. Synthetic trajectories are additionally analyzed by motion type and level. To determine whether some algorithm pairs consistently agree more closely than others, pairwise SSIM and RMSD values are compared across the same 20 trajectories using paired tests. A paired $t$-test is used when normality of the paired differences is not rejected by the Shapiro--Wilk test~\cite{shapiro1965analysis}; otherwise, a Wilcoxon signed-rank test~\cite{wilcoxon1945individual} is used. The statistical significance level is set at $\alpha = 0.05$. Simulations are also visually inspected for realism. 

Finally, the above pipeline is repeated using the motion-free phantom scan to assess whether these discrepancies persist independently of anatomy.

\subsection{Evaluating existing algorithms with APHABAMAS}
\label{sec:methods:accuracy}
APHABAMAS is deployed to assess the absolute simulation accuracy of each algorithm, using the setups in Sec.~\ref{sec:methods:data}. Volume-wise SSIM and RMSD are computed directly against the ground-truth simulation. Unlike the pairwise comparisons in Sec.~\ref{sec:methods:discrepancies}, these metrics provide absolute fidelity measures. For synthetic trajectories, within each motion-type and severity combination, the algorithms are compared using the paired testing procedure described in Sec.~\ref{sec:methods:discrepancies}.

Furthermore, a point-by-point k-space analysis is conducted for the image-based and Type-2 NUFFT-based algorithms by comparing their signals directly against the analytical k-space ground truth. The Type-1 NUFFT-based algorithm is excluded because its non-uniformly spaced signals are mathematically incompatible with point-by-point comparison.

\clearpage
\section{Results}
\label{sec:results}
Sec.~\ref{sec:results:discrepancies} demonstrates the necessity of APHABAMAS, while Sec.~\ref{sec:results:accuracy} reports the accuracy assessment of existing algorithms using APHABAMAS. For the visual results, the figures show representative axial slices for visual clarity while several multi-planar views are provided in Supporting Information.

\subsection{Demonstrating necessity of APHABAMAS}
\label{sec:results:discrepancies}
\begin{figure}
    \centering
    \includegraphics[width=\linewidth]{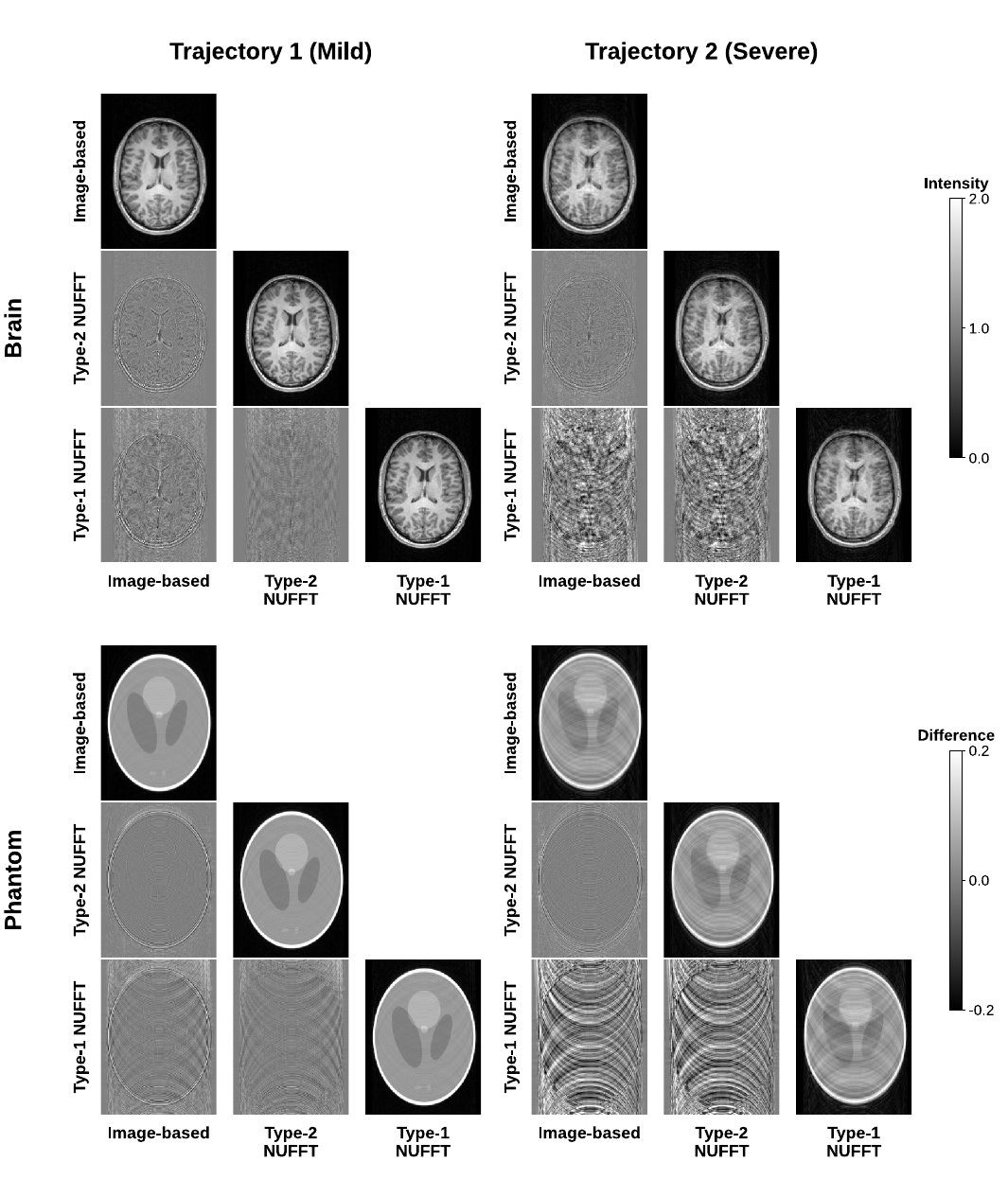}
    \caption{Visual comparison of motion-artifact simulations generated by three algorithms for brain (top) and phantom (bottom) data under tracking-data-derived Trajectories~1 (mild; left) and~2 (severe; right). Within each triangular arrangement, diagonal panels show simulated scans, and off-diagonal panels show pairwise differences computed as the row-labeled algorithm minus the column-labeled algorithm.}
    \label{fig:algorithm_comparisons}
\end{figure}

Fig.~\ref{fig:algorithm_comparisons} visually compares the motion-artifact simulations generated by the three algorithms using brain and phantom images under two tracking-data-derived trajectories. The multi-planar views are shown in Fig.~S4 and S5. While all algorithms produce visually plausible motion artifacts, differences in the simulated artifact patterns are apparent, particularly under severe motion. The Type-1 NUFFT-based simulations exhibit more pronounced differences from the other two algorithms, whereas the Type-2 NUFFT-based and image-based simulations appear more similar. These observations are consistent across the brain and phantom data.

Quantitative pairwise comparisons using SSIM and RMSD across four synthetic motion patterns and three severity levels further confirm these inter-algorithm discrepancies (Fig.~S6 and S7). Statistical tests show significant differences in relative agreement in most comparisons, but which pair agrees more depends on the motion conditions. In general, Type-1 NUFFT-based simulations diverge increasingly from the other methods with increasing motion severity, whereas the agreement between Type-2 NUFFT-based and image-based simulations remains comparatively stable within each motion type. Similar trends in the brain and phantom data support the use of the phantom for assessing algorithmic discrepancies.

Nevertheless, these comparisons only establish differences in relative agreement and they cannot identify which algorithm is more accurate, highlighting the need for ground-truth-enabled assessment using APHABAMAS.

\subsection{Evaluating existing algorithms with APHABAMAS}
\label{sec:results:accuracy}
\begin{figure}
    \centering
    \includegraphics[width=\linewidth]{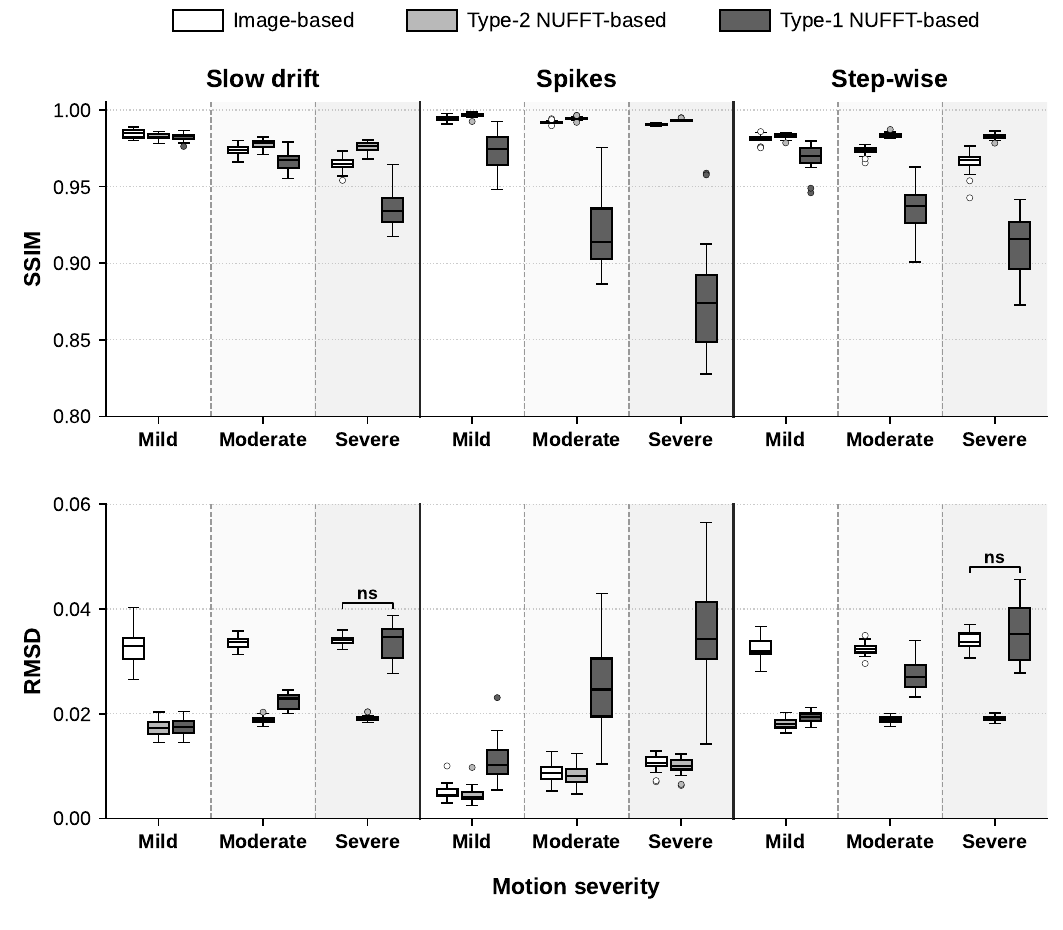}
    \caption{Agreement of the image-based, Type-2 NUFFT-based, and Type-1 NUFFT-based simulation algorithms with the analytical ground truth, measured using SSIM (top row) and RMSD (bottom row). Columns show slow-drift, spikes, and stepwise motion, grouped by mild, moderate, and severe motion. Random-jitter results are presented separately in Fig.~S8 because they require wider axis ranges; this allows tighter limits and clearer visualization of differences in the main figure. Boxes show the interquartile range and median, whiskers extend to the most extreme observations within 1.5 times the interquartile range, and points denote outliers. Higher SSIM and lower RMSD indicate closer agreement with the ground truth. Brackets labeled ``ns'' denote pairwise algorithm comparisons that are not statistically significant after Holm correction. All unmarked within-group algorithm comparisons are statistically significant; $p$-values are reported in Table~S3.}
    \label{fig:different_algorithm_SSIM_RMSD}
\end{figure}

Fig.~~\ref{fig:different_algorithm_SSIM_RMSD} presents the quantitative assessment of simulations using synthetic slow-drift, spike, and stepwise motion trajectories; the corresponding random-jitter results are shown separately in Fig.~~S8 to accommodate their wider metric ranges. The Type-2 NUFFT-based algorithm shows the most consistent agreement with the ground truth, achieving the highest median SSIM and lowest median RMSD in all conditions except SSIM for mild slow-drift motion, for which the image-based method performs marginally better. The advantage of Type-2 becomes more pronounced with increasing motion severity, whereas the accuracy of Type-1 deteriorates substantially. Most pairwise differences between algorithms are significant ($p<0.05$). While RMSD differences between the Type-1 and image-based algorithms are not statistically significant under severe slow-drift ($p=0.516$) or severe stepwise motion ($p=0.163$), Type-1 has substantially lower SSIM in both cases, indicating its poorer accuracy.

\begin{figure}
    \centering
    \includegraphics[width=0.9\linewidth]{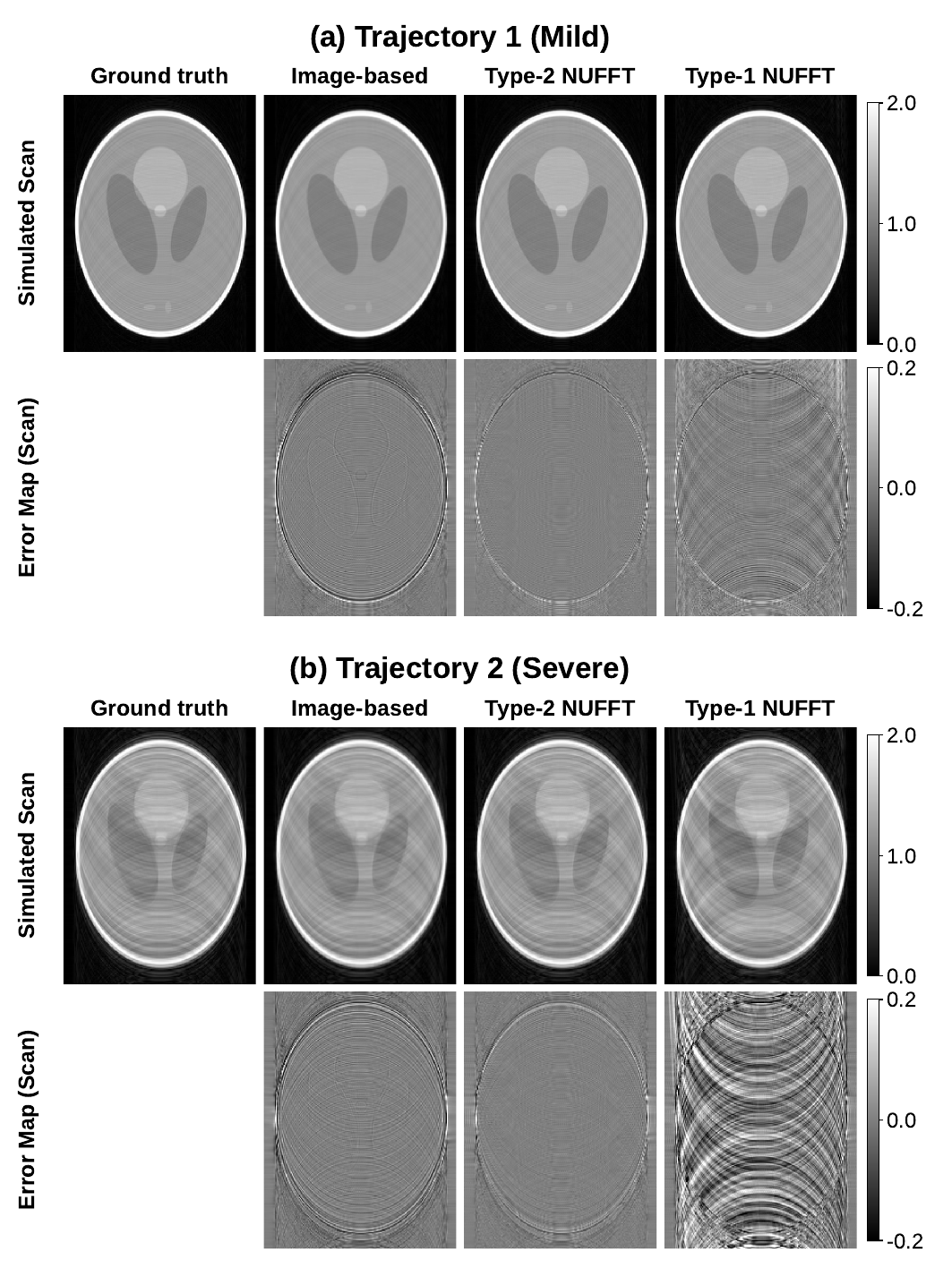}
    \caption{Visualization of simulation results obtained with different algorithms compared to the ground truth for (a) a trajectory of mild motion and (b) a trajectory of severe motion (see Fig.~\ref{fig:motion_trajectories}). In each case, an axial slice of the simulated 3D volume is shown in the first row with corresponding error maps relative to the ground truth in the second.}
    \label{fig:figure10_algorithm_assessment_visual}
\end{figure}

Fig.~\ref{fig:figure10_algorithm_assessment_visual} illustrates the simulations using different algorithms compared against the ground truth under two tracking-data-derived trajectories. For both trajectories, the image-based and Type-2 NUFFT-based algorithms yield simulations visually similar to the ground truth, whereas the Type-1 NUFFT-based algorithm introduces distinct ghosting patterns, as accentuated in the error maps. The image-based algorithm, compared to the Type-2 NUFFT-based algorithm, exhibits increased blurring and slightly larger errors relative to the ground truth. This spatial blurring matches the k-space findings in Fig.~S9, where its signal magnitudes at the k-space edges (high-frequency components) are much lower. The multi-planar views (Fig.~S4 and S5) show that object displacement and pose are broadly similar across algorithms and relative to the ground truth, while the more apparent differences lie in the artifact patterns.

\clearpage
\section{Discussion}
\label{sec:discussion}
Recent DL-based methods for motion detection or correction in high-resolution 3D MRI increasingly rely on simulated motion-corrupted data for their development. This study proposes APHABAMAS to provide an analytical ground truth for assessing how accurately existing simulation algorithms implement the common formulation widely used to generate these data. Utilizing APHABAMAS, we assess three existing algorithms, demonstrating that the Type-2 NUFFT-based algorithm provides the most consistent overall accuracy.

The necessity for APHABAMAS is motivated by two observations from the brain-scan experiments. First, the algorithms yield distinct artifacts from the same inputs, with the magnitude of the discrepancies and the relative agreement between algorithms depending on motion type and severity. Second, despite these differences, all algorithms produced visually plausible artifacts, so visual inspection alone could not establish accuracy. Additionally, we observe that the relative discrepancies between algorithms remain consistent across both brain and phantom scans, supporting the phantom as a surrogate for assessing algorithm behavior on brain data.

APHABAMAS enables quantitative accuracy assessment of the algorithms. Among the evaluated algorithms, the Type-2 NUFFT-based algorithm shows the most consistent agreement with the ground truth across the tested conditions. The image-based algorithm also shows close agreement with the ground truth and achieves marginally higher SSIM under mild slow-drift motion, but generally exhibits larger errors and blurrier simulations than the Type-2 NUFFT-based algorithm. Its performance is expected because it conceptually mimics the MR acquisition process, while the remaining errors likely arise from interpolation during rigid-body transformation of the motion-free scan, as detailed in Sec.~\ref{sec:theory:existing_algorithm}. Moreover, the heavy computational burden of computing many 3D transformations limits its practical utility for simulating complex motion patterns. In contrast, the Type-1 NUFFT-based algorithm generally exhibits distinct artifacts and poorer agreement with the ground truth. Its accuracy also deteriorates substantially with increasing motion severity. This is anticipated, as its formulation contradicts the MR acquisition process. Furthermore, additional investigation reveals that it models translational motion incorrectly when rotational and translational motions occur simultaneously (Fig.~S10); however, the error from this issue is negligible compared to that from violating MR physics (Fig.~S11 and S12) and is thus omitted from our main results.

Our assessment framework is complementary to approaches in the literature that base evaluation on comparison against motion artifacts in real scans, e.g.~\cite{zhao2025accuracies} which reported comparable agreement with acquired motion-corrupted scans for two algorithms, MCP and MDV, corresponding to the Type-1 and Type-2 algorithms we assessed, respectively. However, as they acknowledged, discrepancies between simulated and acquired scans can also arise from motion-tracking errors, and more importantly factors that are not modeled within the formulation that these algorithms adopted. Consequently, comparable agreement with acquired scans does not establish that the two algorithms implement the common underlying formulation with comparable accuracy, which is specifically the question we addressed using APHABAMAS.

APHABAMAS is currently designed for the motion-artifact simulation formulation that is commonly adopted for generating motion-corrupted data for DL training. The flexibility of the analytical phantom may allow extensions to more complex simulation formulations, such as those that accommodate additional k-space sampling patterns or account for multi-coil acquisition effects. For instance, when raw MR signals are accessible~\cite{johnson2019conditional, nghiem2026network, polak2022scout}, developing motion correction methods may require simulations incorporating coil sensitivities and parallel imaging, which can alter artifact appearance~\cite{tianqi2024, zhao2025accuracies}. However, there are motion-induced effects, such as changes in local $B_0$ and $B_1$, that may not be assessed with APHABAMAS, as it is unclear if it is possible to incorporate these while retaining a closed-form reference. Finally, an important question for the future is how the accuracy of simulated training data affects downstream DL-based motion detection or correction.

\section{Conclusion}
\label{sec:conclusion}
APHABAMAS provides a rigorous tool for assessing how accurately existing algorithms implement the common formulation for simulating motion artifacts in high-resolution 3D MRI, a formulation widely used to generate motion-corrupted data for training DL models. Of the three algorithms assessed, the Type-2 NUFFT-based algorithm agreed most consistently with the analytical ground truth across the tested conditions.

\clearpage
\section*{Acknowledgments}
This work is supported by the EPSRC-funded UCL Centre for Doctoral Training in Intelligent, Integrated Imaging in Healthcare (i4health)
[EP/S021930/1].

\section*{Funding}
This work is supported by Engineering and Physical Sciences Research Council, [EP/S021930/1].

\section*{Data Availability Statement}
Code for all experiments and figure generation is publicly available at: \url{https://github.com/zcahtwu/APHABAMAS}.

\bibliographystyle{MRM-AMA} 
\bibliography{reference.bib} 

\clearpage
\include{SI}

\end{document}

%% file: SI.tex
\section*{Supporting Information}

\setcounter{subsection}{0}
\renewcommand{\thesubsection}{S\arabic{subsection}}

\setcounter{subsubsection}{0}
\renewcommand{\thesubsubsection}
{\thesubsection.\arabic{subsubsection}}

\setcounter{equation}{0}
\renewcommand{\theequation}{S\arabic{equation}}

\subsection{Additional derivations and descriptions}
\label{sec:si:text}

\subsubsection{How to specify \texorpdfstring{$\rho_t(\vec{r})$}{rho\_t(r)} in terms of \texorpdfstring{$\rho_0(\vec{r})$}{rho\_0(r)} and \texorpdfstring{$A_t(\vec{r})$}{A\_t(r)}}
Here we present how $\rho_t(\vec{r})$ is expressed in terms of $\rho_0(\vec{r})$ and $A_t(\vec{r})$, with the illustration shown below. If we denote some location on the object at time 0 as $\vec{r}_0$, the amount of spin contained within the infinitesimal space centered at that location is $\rho_0(\vec{r}_0)d\vec{r}_0$. The corresponding location at time $t$, denoted as $\vec{r}_t$, can be determined with $A_t(\vec{r})$,
\begin{align}
    \vec{r}_t = A_t(\vec{r}_0) \, ,
    \label{eq:A_t}
\end{align}
and the corresponding infinitesimal space is 
\begin{align}
    d\vec{r}_t= |J_t(\vec{r}_0)|d\vec{r}_0 \, ,
    \label{eq:correspond_space}
\end{align}
where $J_t(\vec{r})$ denotes the Jacobian matrix of $A_t(\vec{r})$ and $|J_t(\vec{r})|$ the Jacobian determinant. If we assume local conservation of spins:
\begin{align}
    \rho_0(\vec{r}_0) \, d\vec{r}_0 = \rho_t(\vec{r}_t) \, d\vec{r}_t \, ,
    \label{eq:spin_conservation}
\end{align} 
$\rho_t(\vec{r})$ can be determined from $\rho_0(\vec{r})$ and $A_t(\vec{r})$. 
\begin{figure}[h]
    \centering
    \includegraphics[width=\linewidth]{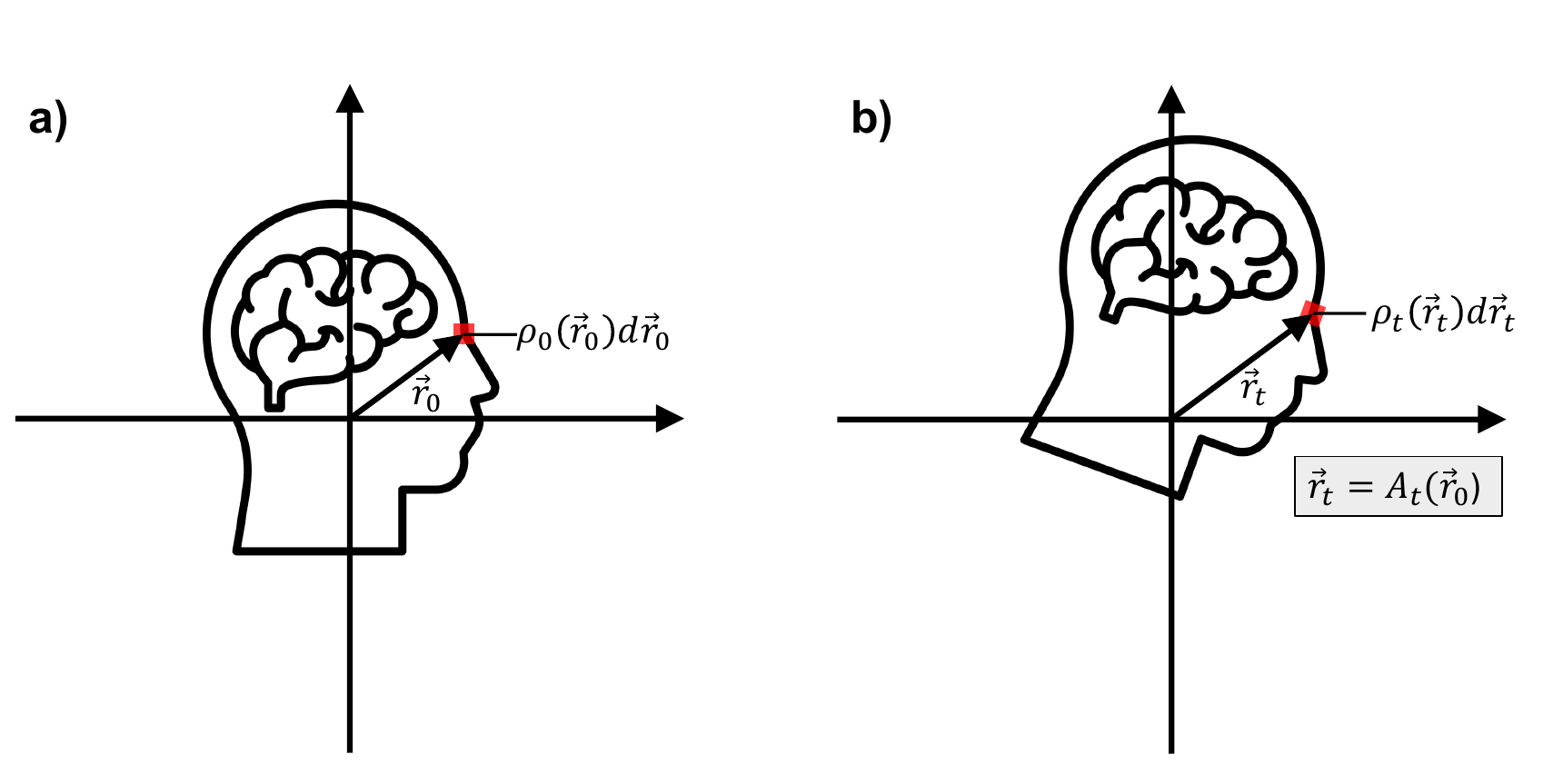}
    \caption*{\textbf{Derivation schematic.}
    Schematic 2D illustration for defining $\rho_t(\vec{r})$
    using $\rho_0(\vec{r})$ and $A_t(\vec{r})$. The object at
    time $t=0$, with spin density $\rho_0(\vec{r})$, is shown
    in (a), while (b) shows the object at time $t$ after applying
    $A_t(\vec{r})$, with spin density $\rho_t(\vec{r})$. A point
    at $\vec{r}_0$ at time $0$ is mapped to
    $\vec{r}_t=A_t(\vec{r}_0)$ at time $t$.}
\end{figure}

\noindent Applying substitutions using Eq.~\eqref{eq:A_t} and Eq.~\eqref{eq:correspond_space}, we obtain 
\begin{align}
\rho_t(A_t(\vec{r}_0)) = \frac{\rho_0(\vec{r}_0)}{|J_t(\vec{r}_0)|} {\, .}
\label{eq:density_relations}
\end{align}
Under the common assumption that head motion is rigid, local conservation of spins is satisfied. Additionally, we have $|J_t(\vec{r}_0)| = 1$, so Eq.~\eqref{eq:density_relations} simplifies to $\rho_t(A_t(\vec{r}_0)) = \rho_0(\vec{r}_0)$.
This implies that $\rho_t(\vec{r})$ can be expressed with $\rho_0(\vec{r})$ and $A_t(\vec{r})$ as
\begin{align}
\rho_t(\vec{r}) = \rho_0(A_t^{-1}(\vec{r})) {\, ,}
\end{align}
where $A_t^{-1}$ denotes the inverse of $A_t$.

\subsubsection{Construction of the rotation matrix}
Here we specify how the three rotational parameters of the motion trajectory, $\theta_{x,t}$, $\theta_{y,t}$, and $\theta_{z,t}$, are combined to construct the rotation matrix $\mathbf{R}_t$. The parameters specify rotations about the $x$-, $y$-, and $z$-axes, respectively. The corresponding axis-specific
rotation matrices are

\begin{align}
\mathbf{R}_x(\theta_{x,t}) =
\begin{bmatrix}
1 & 0 & 0 \\
0 & \cos\theta_{x,t} & -\sin\theta_{x,t} \\
0 & \sin\theta_{x,t} & \cos\theta_{x,t}
\end{bmatrix},
\label{eq:rotation_x}
\end{align}
\begin{align}
\mathbf{R}_y(\theta_{y,t}) =
\begin{bmatrix}
\cos\theta_{y,t} & 0 & \sin\theta_{y,t} \\
0 & 1 & 0 \\
-\sin\theta_{y,t} & 0 & \cos\theta_{y,t}
\end{bmatrix},
\label{eq:rotation_y}
\end{align}
\begin{align}
\mathbf{R}_z(\theta_{z,t}) =
\begin{bmatrix}
\cos\theta_{z,t} & -\sin\theta_{z,t} & 0 \\
\sin\theta_{z,t} & \cos\theta_{z,t} & 0 \\
0 & 0 & 1
\end{bmatrix}.
\label{eq:rotation_z}
\end{align}
The composite rotation matrix used in this work is
\begin{align}
\mathbf{R}_t =
\mathbf{R}_z(\theta_{z,t})
\mathbf{R}_x(\theta_{x,t})
\mathbf{R}_y(\theta_{y,t}).
\label{eq:combined_rotation}
\end{align}
These are defined in a right-handed coordinate system, with positive rotations following the right-hand rule.

\subsubsection{Type-2 NUDFT as a k-space interpolation}
Here, we explain that approximating the k-space signal at arbitrary k-space locations from a motion-free scan using a Type-2 NUDFT can be seen as an interpolation in k-space. For clarity and simplicity, we present this derivation in 1D.

Assume we have access to the complex-valued motion-free scan, $\hat{\rho}(x)$, reconstructed from $N$ uniformly sampled k-space signals, $S(k_m)$. Here $k_m = m \Delta k$ is the sampling location, where $m \in [-N/2, N/2 - 1]$ is an integer, and $\Delta k$ is the sampling interval. The complex-valued scan at spatial locations $x_n = n \Delta x$ (where $n \in [-N/2, N/2 - 1]$ and $\Delta x \Delta k = 1/N$) is defined by the IDFT of the sampled signals:
\begin{align}
\hat{\rho}(x_n) = \frac{1}{N} \sum_{m=-N/2}^{N/2-1} S(k_m) e^{i 2\pi k_m x_n} \label{eq:idft}
\end{align}
By definition, the Type-2 NUDFT applied to $\hat{\rho}(x_n)$ to approximate the signal at an arbitrary k-space location $k$ is given by:
\begin{align}
S_{\text{NUDFT}}(k) = \sum_{n=-N/2}^{N/2-1} \hat{\rho}(x_n) e^{-i 2\pi k x_n} \label{eq:nudft}
\end{align}
By substituting Eq.~\eqref{eq:idft} into Eq.~\eqref{eq:nudft}:
\begin{align}
S_{\text{NUDFT}}(k) = \sum_{n=-N/2}^{N/2-1} \left( \frac{1}{N} \sum_{m=-N/2}^{N/2-1} S(k_m) e^{i 2\pi k_m x_n} \right) e^{-i 2\pi k x_n}
\end{align}
By reversing the order of summation, we can isolate the terms dependent on $n$ and group the complex exponentials:
\begin{align}
S_{\text{NUDFT}}(k) = \sum_{m=-N/2}^{N/2-1} S(k_m) \left[ \frac{1}{N} \sum_{n=-N/2}^{N/2-1} e^{-i 2\pi x_n (k - k_m)} \right] \label{eq:plug_idft_to_nudft}
\end{align}
Recalling the DFT relationship $x_n = n \Delta x = \frac{n}{N \Delta k}$, the exponent can be rewritten as $-i \frac{2\pi}{N} n \left( \frac{k - k_m}{\Delta k} \right)$. The inner summation is a finite geometric series of complex exponentials, which allows us to simplify Eq.~\eqref{eq:plug_idft_to_nudft} to the following:
\begin{align}
S_{\text{NUDFT}}(k) = \sum_{m=-N/2}^{N/2-1} S(k_m) \left[ \frac{1}{N} e^{i \frac{\pi}{N} \left( \frac{k - k_m}{\Delta k} \right)} \frac{\sin\left( \pi \left( \frac{k - k_m}{\Delta k} \right) \right)}{\sin\left( \frac{\pi}{N} \left( \frac{k - k_m}{\Delta k} \right) \right)} \right] \label{eq:final_interpolation_eq}
\end{align}
Eq.~\eqref{eq:final_interpolation_eq} shows that applying a Type-2 NUDFT to the complex-valued motion-free scan is equivalent to interpolating the original k-space samples $S(k_m)$. It also shows that $S_{\text{NUDFT}}(k)$ is periodic.

When the magnitude image is used as input, as in our experiments, the same derivation applies after replacing $S(k_m)$ with the DFT coefficients of that magnitude image. Thus, the k-space interpolation interpretation remains valid, although these coefficients need not coincide with the original sampled signals. Taking the magnitude image as input will in general introduce an additional discrepancy relative to those signals, beyond the interpolation error described above.

\subsubsection{Generation of the synthetic motion trajectories}
Each trajectory comprised six degrees of freedom, with translations in millimeters and Euler rotations in degrees. Time was normalized to the acquisition duration, and all trajectories began at the zero-motion reference pose. Twenty reproducible trajectories were generated for each combination of motion type and severity level. The implementations are available in our \href{https://github.com/zcahtwu/APHABAMAS}{[GitHub repository]}. The motion severities for each type of motion are listed in the table below as well.

\noindent\textbf{Slow drift:}\\
Slow drift represented gradual, continuously varying changes in head pose over the acquisition. Each trajectory was constructed from 128 control poses with time-varying directions and relative contributions. Component weighting produced either nodding-dominant motion or head-shaking-dominant motion. Temporal smoothing is applied to avoid sudden jitters. Translation and rotation were scaled independently so that their maximum Euclidean magnitudes matched the selected motion level. Slow drift was therefore controlled by its largest translation and rotation.\\

\noindent\textbf{Spikes:}\\
Spike motion represented brief, isolated changes in pose followed by an immediate return to the reference pose. The central part of the acquisition was divided into temporal sections, with one event placed within each section to distribute the events over time. Each event lasted one of the 256 acquisition intervals (one TR) and contained either nodding-dominant or head-shaking-dominant motion; the other components remained zero. Event direction varied, and amplitude was restricted to 70--100\% of the limit for the selected level. Spike motion was controlled by both event count and maximum translation and rotation.\\

\noindent\textbf{Stepwise motion:}\\
Stepwise motion represented abrupt changes in head pose that were maintained until the next movement. Each new motion change was added to the preceding pose, producing a sequence of distinct states. Changes alternated between nodding-dominant and head-shaking-dominant motion, with smaller contributions from the remaining degrees of freedom. For $N$ movements, the acquisition contained $N+1$ constant-pose states (1 motion-free). Their timings varied between trajectories, with no state occupying more than one third of the acquisition. The complete trajectory was scaled independently in translation and rotation to reach the selected level. Stepwise motion was controlled by both the number of persistent pose changes and the maximum displacement from the initial reference position.\\

\noindent\textbf{Random jitter:}\\
Random jitter represented rapid, irregular fluctuations in pose throughout the acquisition. Each trajectory contained 256 equally spaced intervals, with the pose held constant within each interval. The first pose was zero. For the remaining 255 intervals, all six pose components were initially drawn independently from a standard normal distribution. Translation and rotation were then scaled independently across the complete trajectory so that their maximum Euclidean magnitudes reached the selected severity limits.\\

\noindent\textbf{Motion severity definitions:}\\
\begin{center}
\small
\begin{tabular}{llccc}
\hline
Motion type & Parameter & Mild & Moderate & Severe \\
\hline
Slow drift
& Peak translation/rotation
& $1\,\mathrm{mm}/1^\circ$
& $3\,\mathrm{mm}/3^\circ$
& $5\,\mathrm{mm}/5^\circ$ \\

Spike motion
& Number of events
& $2-3$ & $4-5$ & $6-7$ \\
& Translation/rotation limits
& $1\,\mathrm{mm}/1^\circ$
& $3\,\mathrm{mm}/3^\circ$
& $5\,\mathrm{mm}/5^\circ$ \\

Stepwise motion
& Number of movements
& $2-3$ & $4-5$ & $6-7$ \\
& Peak translation/rotation
& $1\,\mathrm{mm}/1^\circ$
& $3\,\mathrm{mm}/3^\circ$
& $5\,\mathrm{mm}/5^\circ$ \\

Random jitter
& Peak translation/rotation
& $0.5\,\mathrm{mm}/0.5^\circ$
& $1.0\,\mathrm{mm}/1.0^\circ$
& $1.5\,\mathrm{mm}/1.5^\circ$ \\
\hline
\end{tabular}
\end{center}
Translation and rotation magnitudes denote the Euclidean norms of the three-component translation and Euler-angle vectors, respectively. The rotation magnitude is therefore a norm of the Euler-angle parameters, rather than the equivalent axis--angle rotation. Slow drift, stepwise motion, and random jitter were scaled to reach the tabulated peak values. For spike motion, the tabulated values were upper limits, with individual event amplitudes ranging from 70--100\% of those limits.

Smaller amplitude limits were used for random jitter to represent small, rapid pose fluctuations. Jitter changed pose at every acquisition interval, whereas slow drift varied smoothly and spike and stepwise motion contained relatively few abrupt events. Consequently, matching the maximum translation and rotation across these models would not match their temporal variability or the changes between successive poses. The severity levels were therefore defined within each motion type and should not be interpreted as equivalent levels of image degradation across motion types.

\clearpage
\subsubsection{Derivation of the Cartesian k-space sampling locations}
\label{sec:si-kspace-sampling}

The Cartesian k-space grid for reconstructing the motion-free image is derived from the image matrix size and voxel dimensions. Let $(N_x,N_y,N_z)$ denote the image matrix size and
$(\Delta x,\Delta y,\Delta z)$ the voxel dimensions in millimeters.
The field of view along each axis is
\begin{equation}
    L_x=N_x\Delta x,\qquad
    L_y=N_y\Delta y,\qquad
    L_z=N_z\Delta z.
\end{equation}
Using the MRI Fourier convention with kernel
$\exp(-i2\pi\mathbf{k}^{\mathsf T}\mathbf{r})$, k-space coordinates
are expressed in 1/mm, and their sampling intervals are
\begin{equation}
    \Delta k_x=\frac{1}{L_x},\qquad
    \Delta k_y=\frac{1}{L_y},\qquad
    \Delta k_z=\frac{1}{L_z}.
\end{equation}
For the even matrix dimensions used in this study, the centered
coordinate lists are
\begin{equation}
    k_{a,j}=\left(j-\frac{N_a}{2}\right)\Delta k_a,
    \qquad j=0,\ldots,N_a-1,\quad a\in\{x,y,z\}.
\end{equation}
The list of sampling locations comprises all Cartesian
combinations $(k_{x,j},k_{y,\ell},k_{z,m})$, giving
$N_xN_yN_z$ locations at which the analytical Fourier transform
of the phantom is evaluated.
\clearpage
\subsection{Additional figures and tables}
\label{sec:si:figures_tables}

\setcounter{figure}{0}
\renewcommand{\thefigure}{S\arabic{figure}}

\setcounter{table}{0}
\renewcommand{\thetable}{S\arabic{table}}

\begin{figure}[h]
    \centering
    \includegraphics[width=0.8\linewidth]{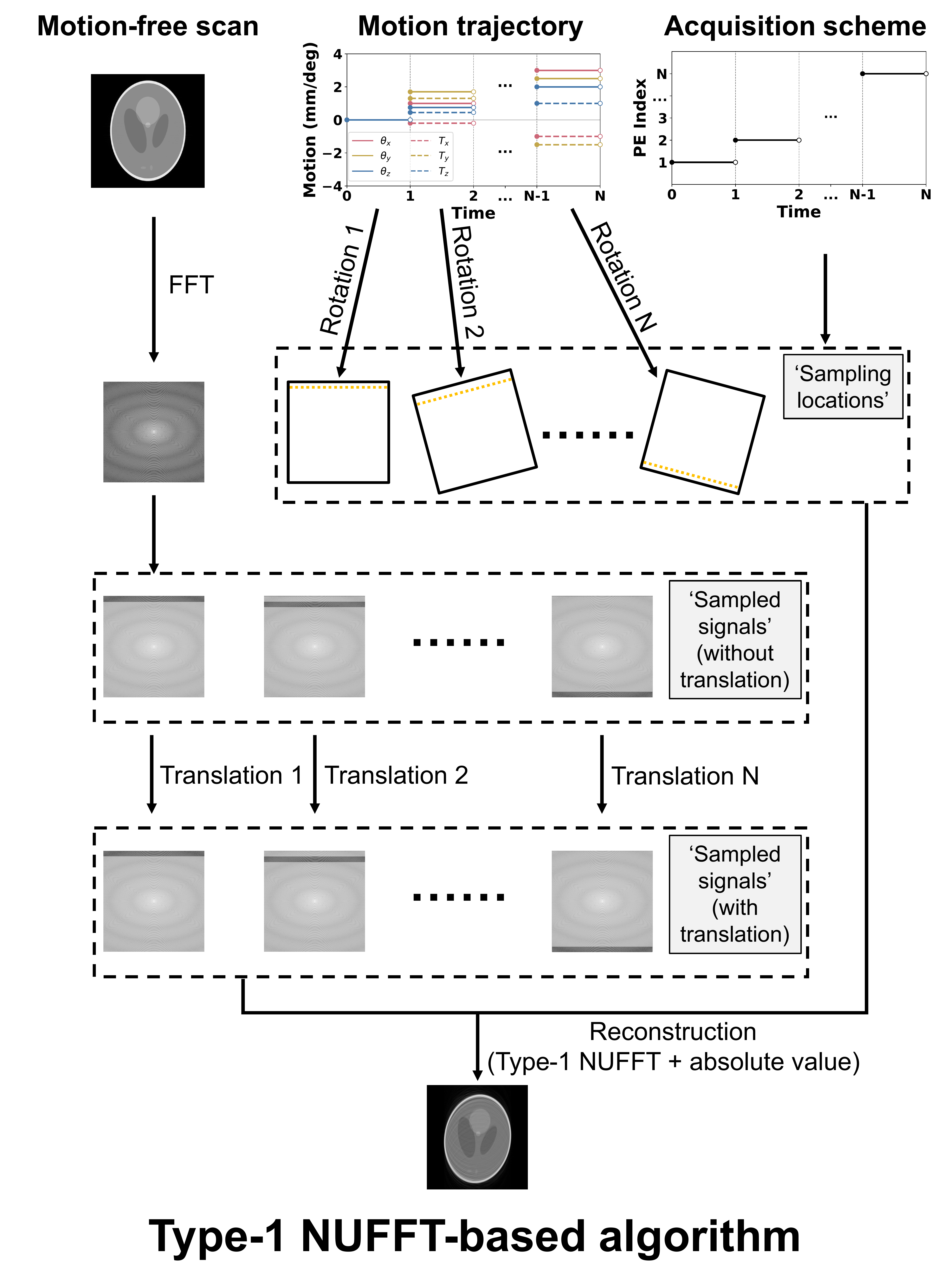}
    \caption{Visualization of the Type-1 NUFFT-based algorithm.}
    \label{figS:type1_visualization}
\end{figure}

\clearpage
\clearpage
\begin{sidewaysfigure}[p]
    \centering
    \includegraphics[width=\linewidth]{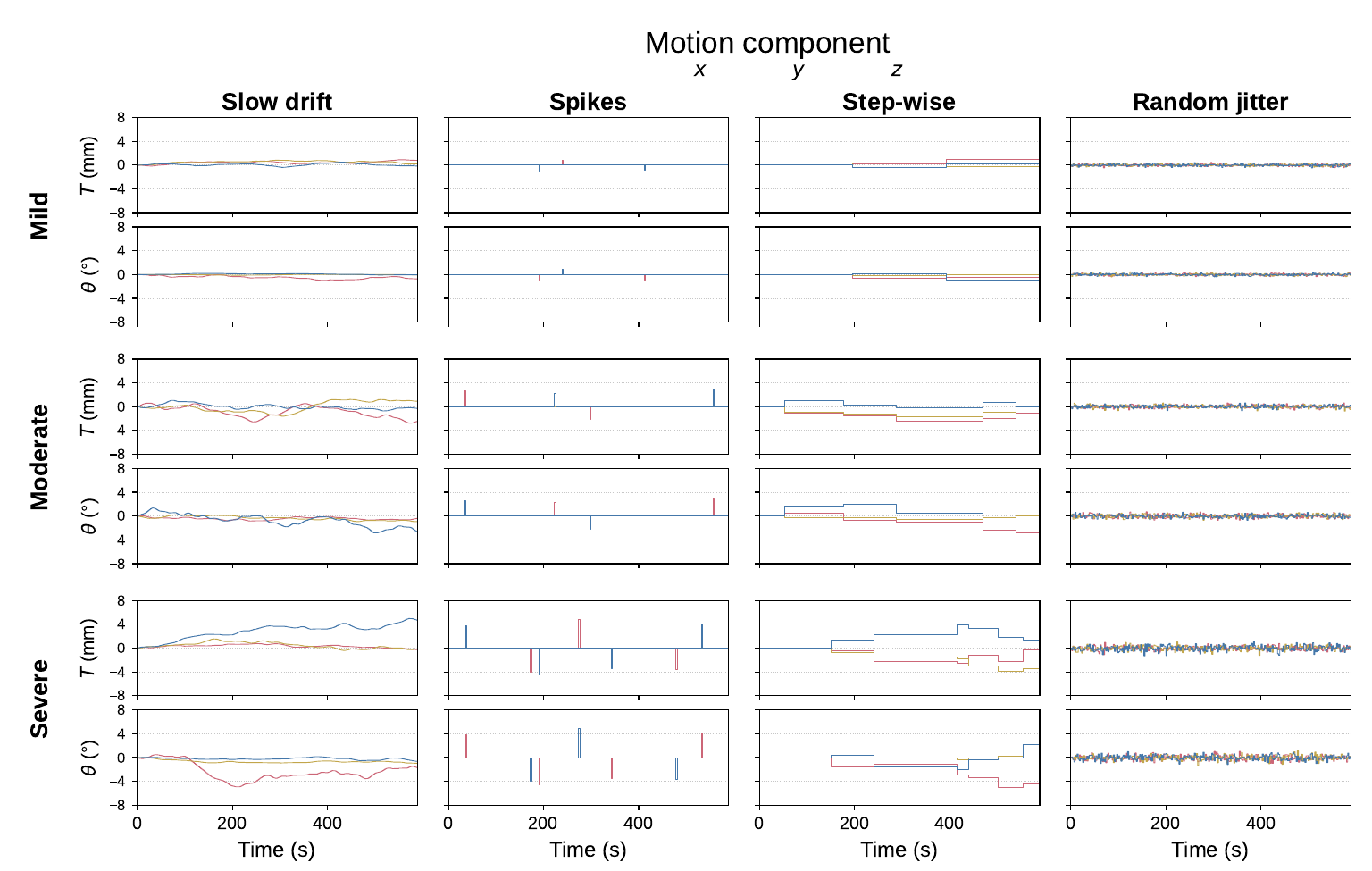}
    \caption{Examples of synthetic motion trajectories of four motion patterns at three motion severity levels.}
    \label{fig:synthetic_trajectories}
\end{sidewaysfigure}
\clearpage

\begin{figure}[p]
    \centering
    \includegraphics[width=\linewidth]{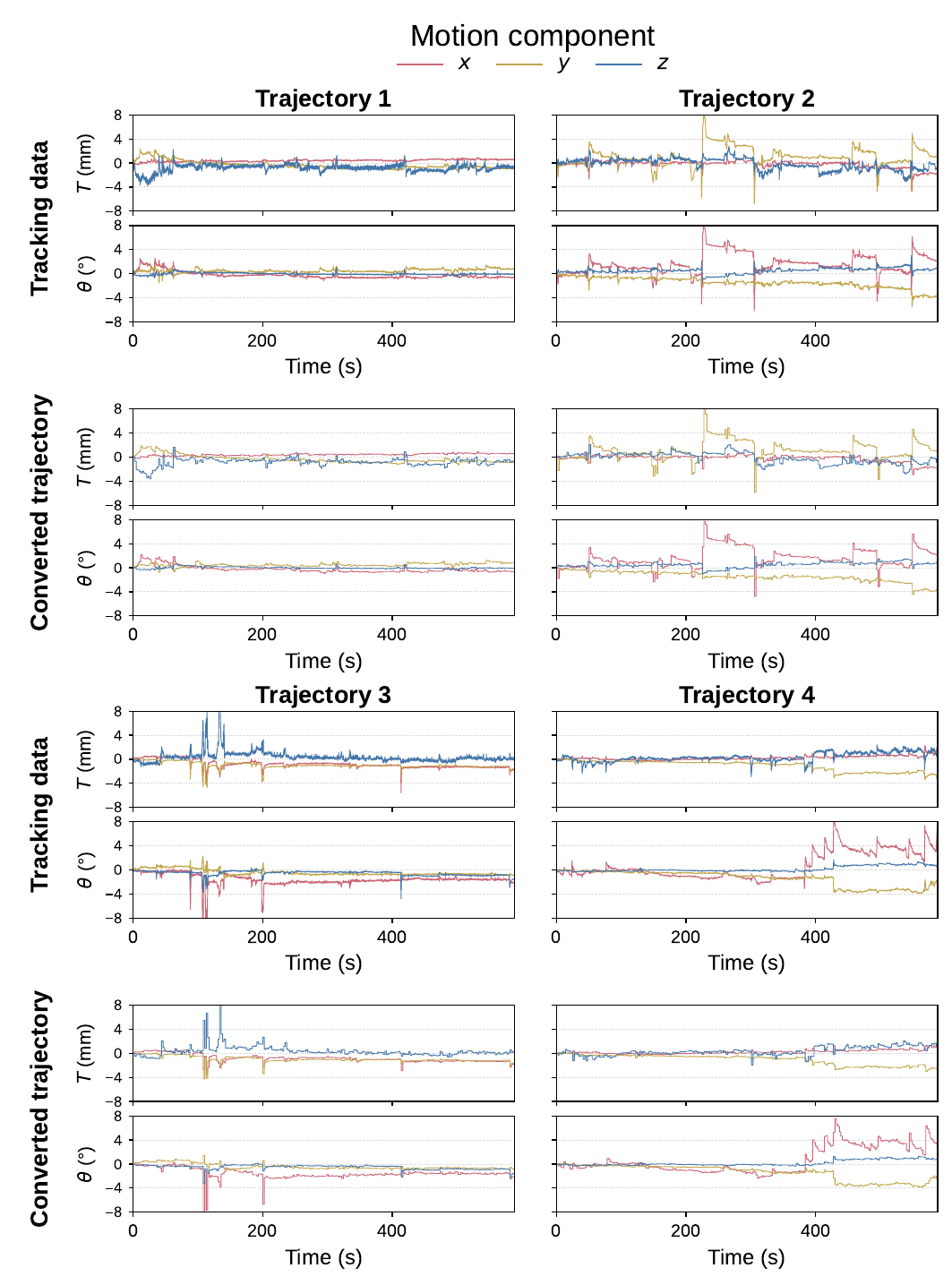}

    \caption{All eight sets of motion-tracking data and their
    corresponding converted motion trajectories.}
    \label{figS:tracking_trajectories}
\end{figure}

\clearpage

\begin{figure}[p]
    \ContinuedFloat
    \centering
    \includegraphics[width=\linewidth]{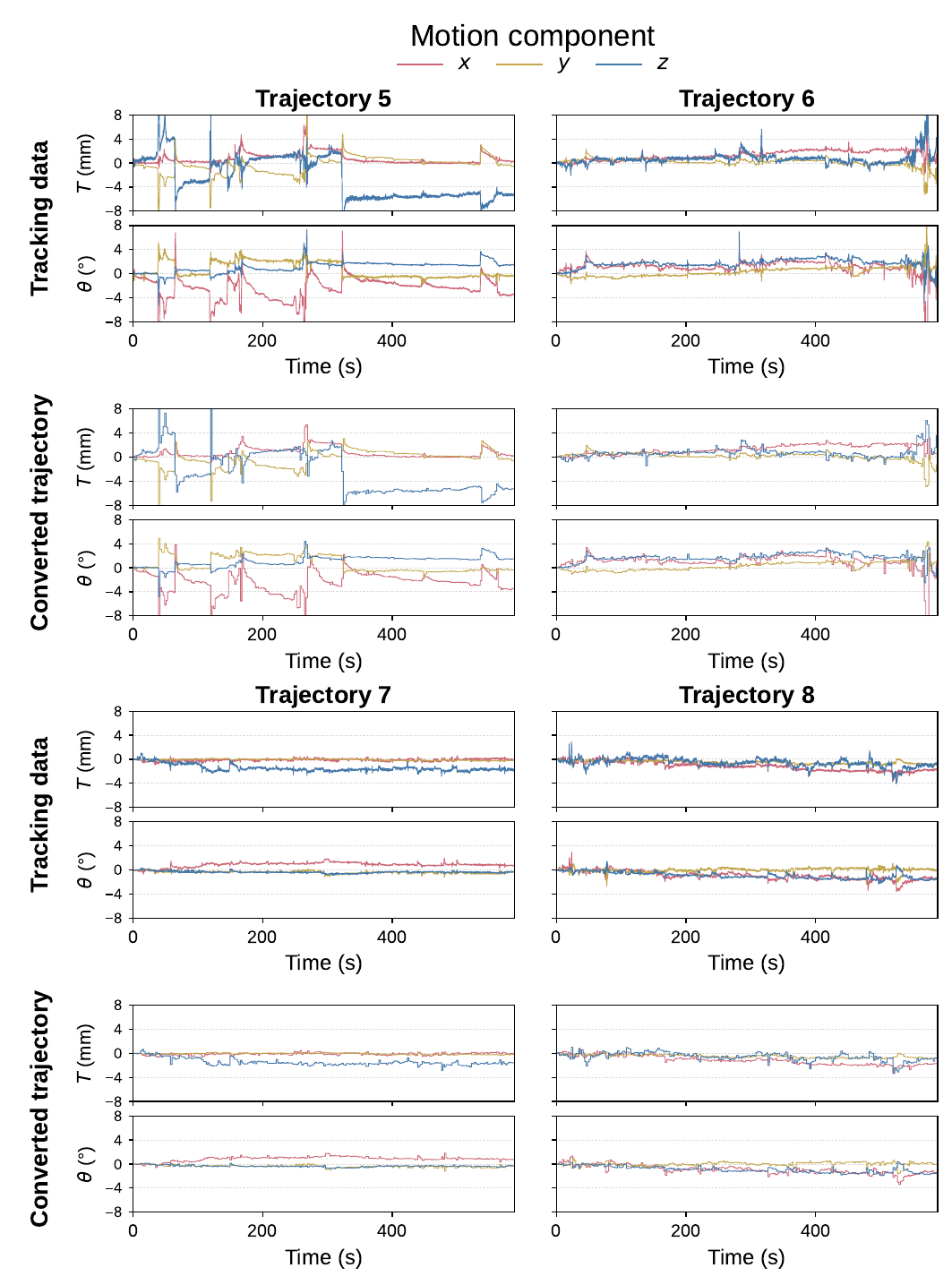}

    \caption[]{All eight sets of motion-tracking data and their
    corresponding converted motion trajectories (continued).}
\end{figure}

\clearpage

\begin{figure}[p]
    \centering
    \includegraphics[width=\linewidth]{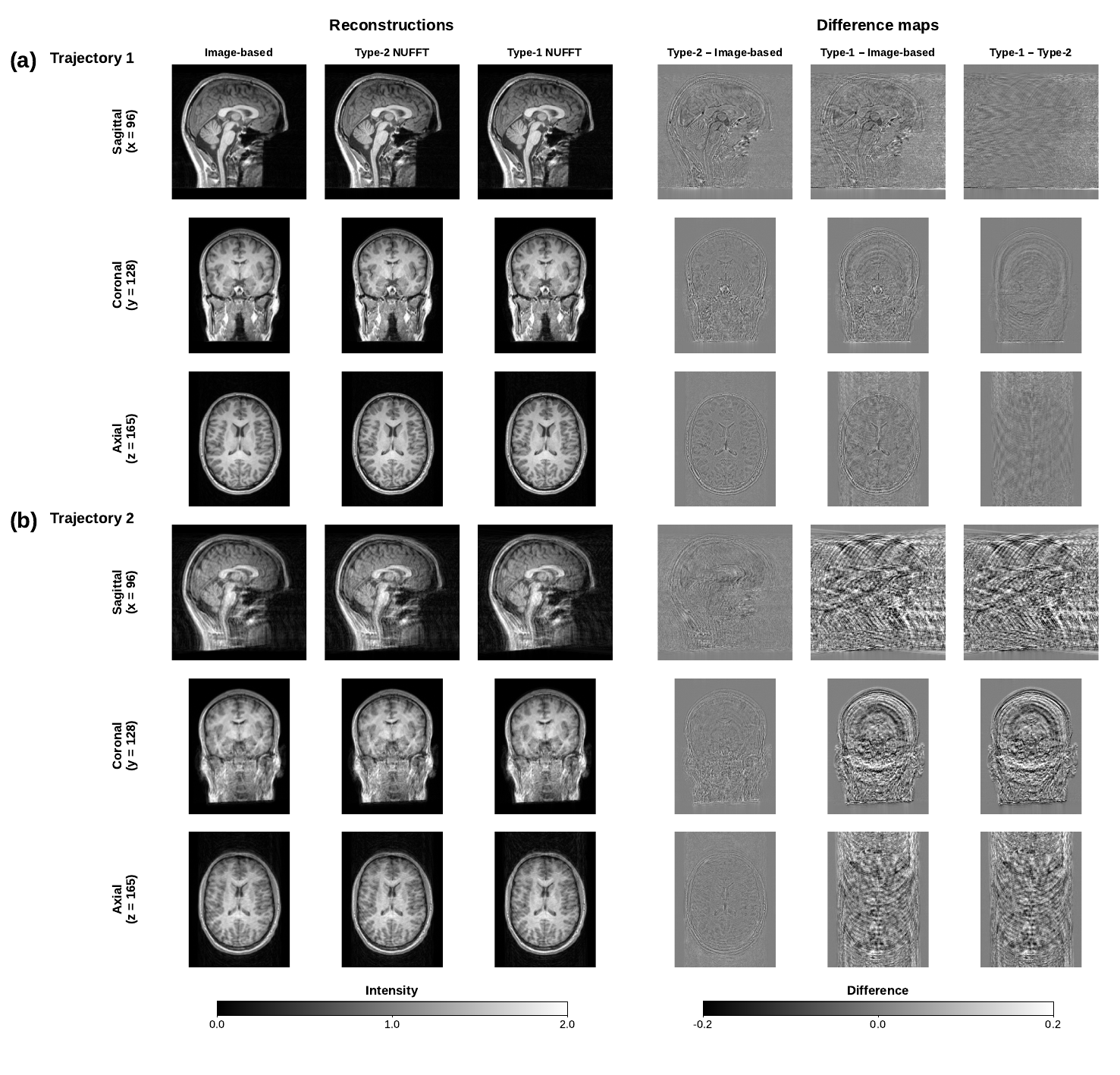}

    \caption{Multi-planar views of the brain simulations for (a) the relatively mild Trajectory 1 and (b) the more severe Trajectory 2 from Fig.~7. For each trajectory, rows show sagittal, coronal, and axial views. The first three columns show simulations generated using the image-based, Type-2 NUFFT-based, and Type-1 NUFFT-based algorithms, respectively. The remaining columns show the indicated pairwise difference maps. Object displacement and pose appear broadly similar across algorithms, while the more apparent differences occur in the simulated artifact patterns.}
    \label{figS:brain_multi_planar}
\end{figure}

\begin{figure}[p]
    \centering
    \includegraphics[width=\linewidth]{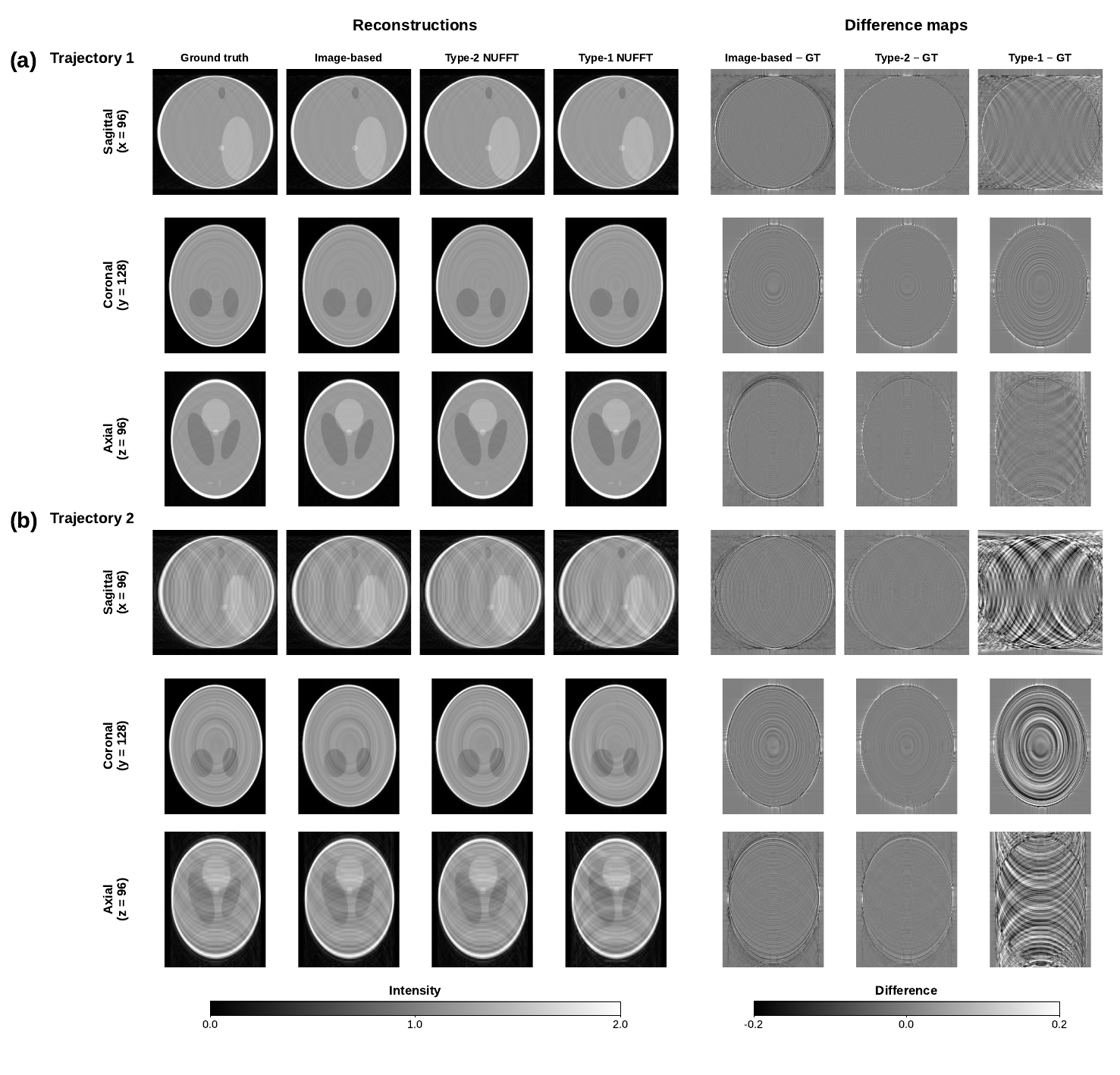}

    \caption{Multi-planar views of the phantom simulations for (a) the relatively mild Trajectory 1 and (b) the more severe Trajectory 2 from Fig.~7. For each trajectory, rows show sagittal, coronal, and axial views. The reconstruction columns show the analytical ground truth followed by simulations generated using the image-based, Type-2 NUFFT-based, and Type-1 NUFFT-based algorithms. The remaining columns show the difference maps of each algorithm relative to the ground truth. Object displacement and pose are broadly consistent across algorithms and with the ground truth, while the more apparent differences occur in the simulated artifact patterns.}
    \label{figS:phantom_multi_planar}
\end{figure}

\clearpage
\begin{sidewaysfigure}[p]
    \centering
    \includegraphics[width=0.8\textheight]{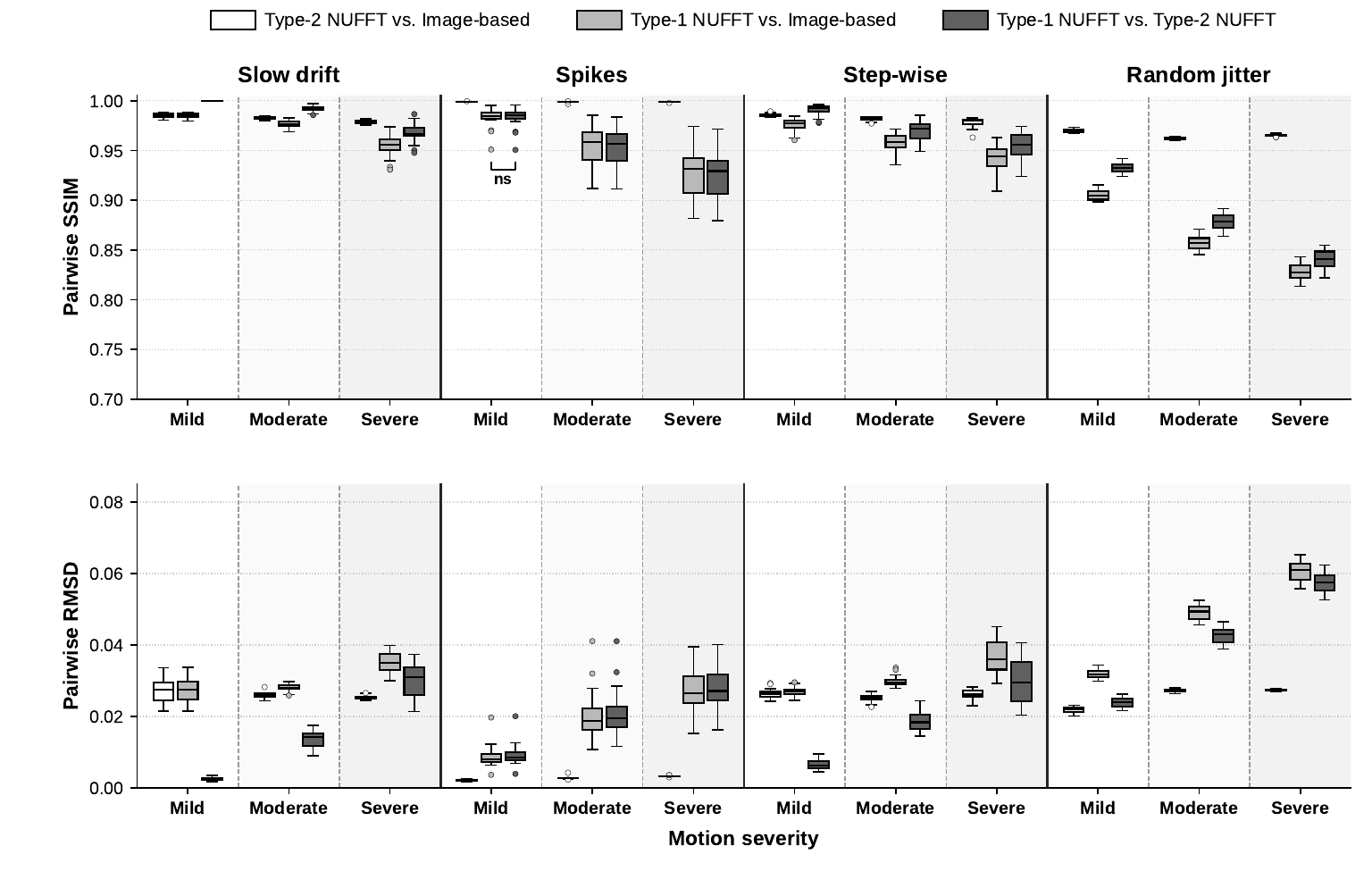}
    \caption[Pairwise algorithm comparisons for brain data]{
    Pairwise agreement between motion-simulation algorithms for brain data.
    SSIM (top row) and RMSD (bottom row) are shown for slow drift,
    spike motion, stepwise motion, and random jitter at mild, moderate,
    and severe motion levels. Each boxplot summarizes 20 trajectories.
    Higher SSIM and lower RMSD indicate closer agreement between the
    corresponding algorithms. Boxes show the interquartile range,
    horizontal lines indicate medians, whiskers extend to the most
    extreme observations within 1.5 times the interquartile range,
    and points indicate outliers.
    Brackets labelled ``ns'' identify comparisons between the plotted
    algorithm-pair groups that were not statistically significant
    after Holm correction ($p_{\mathrm{Holm}} \geq 0.05$). Statistical comparisons are reported in Table S1.
    }
    \label{fig:si-brain-pairwise}
\end{sidewaysfigure}

\begin{sidewaysfigure}[p]
    \centering
    \includegraphics[width=0.8\textheight]{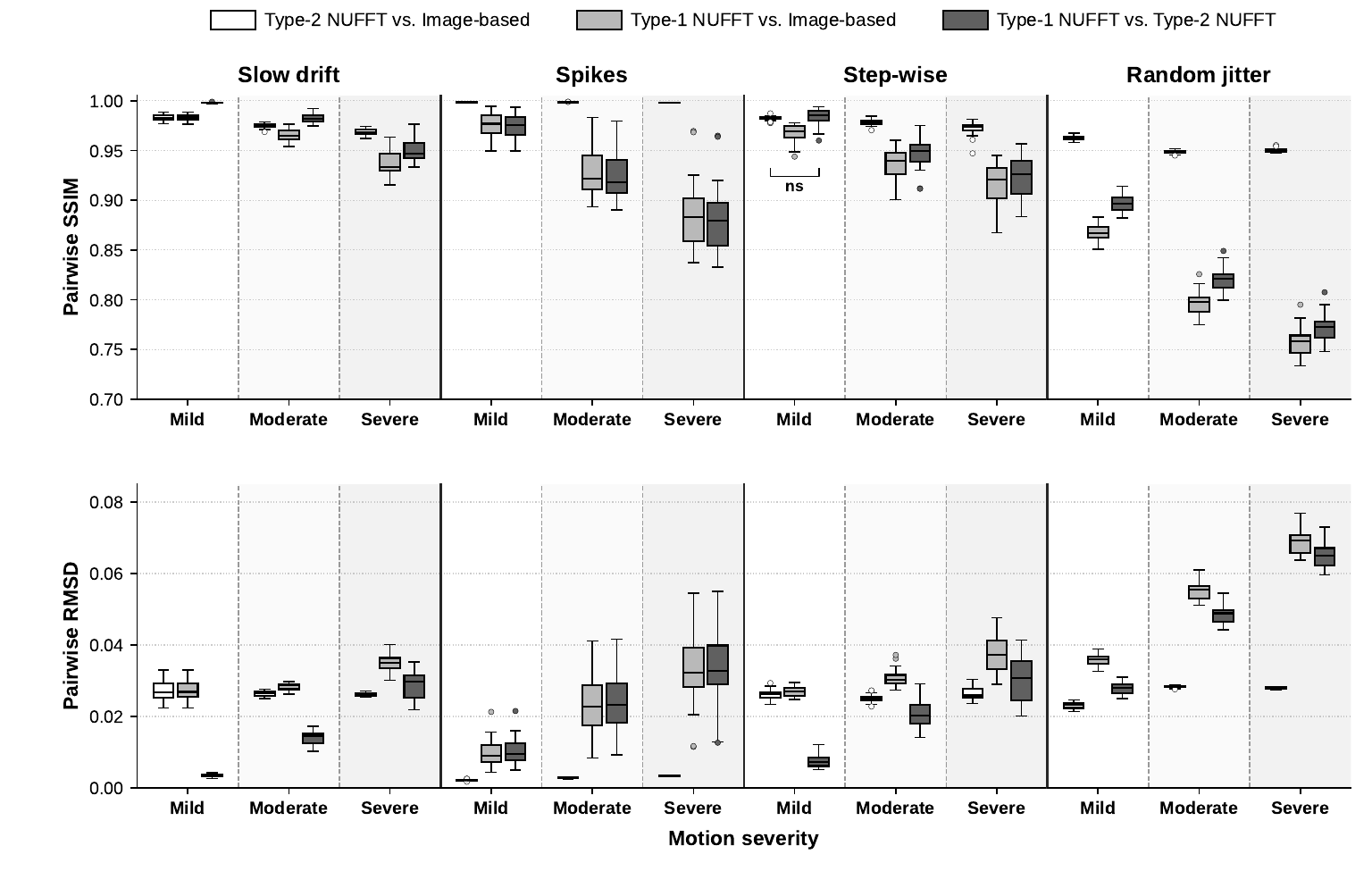}
    \caption[Pairwise algorithm comparisons for the digital phantom]{
    Pairwise agreement between motion-simulation algorithms for the
    digital phantom. SSIM (top row) and RMSD (bottom row) are shown
    for slow drift, spike motion, stepwise motion, and random jitter
    at mild, moderate, and severe motion levels.
    Each boxplot summarizes 20 trajectories, using the same plotting
    conventions and statistical annotations as
    Fig.~\ref{fig:si-brain-pairwise}.
    These metrics compare algorithms with one another;
    they do not measure accuracy relative to analytical ground truth. Statistical comparisons are reported in Table S2.
    }
    \label{fig:si-phantom-pairwise}
\end{sidewaysfigure}

\begin{figure}[p]
    \centering
    \includegraphics[width=\linewidth]{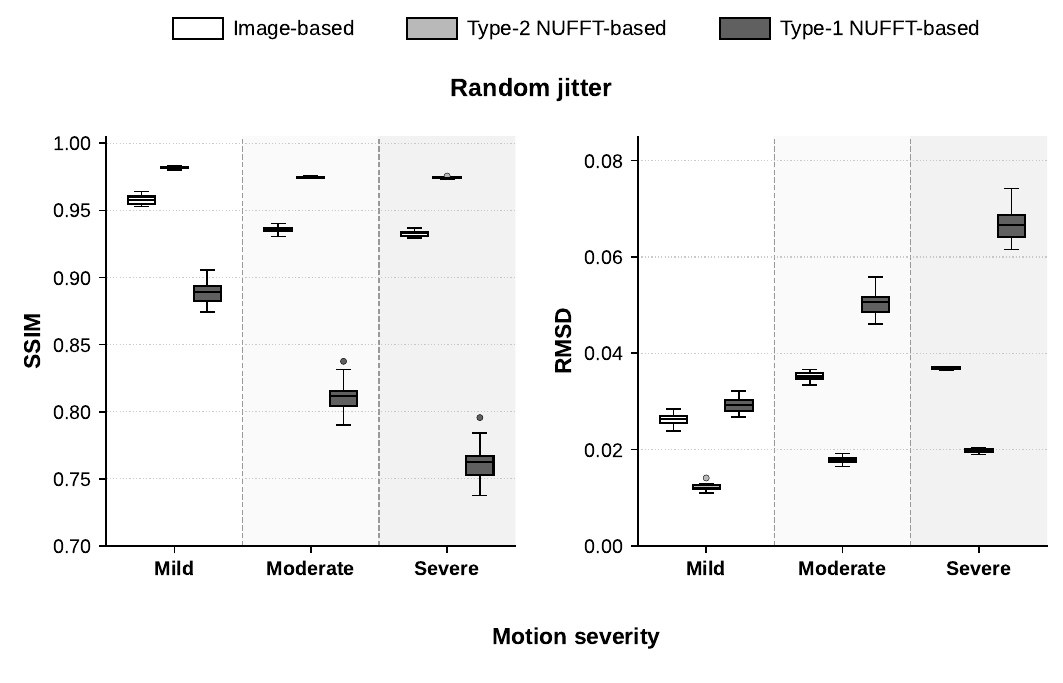}
    \caption{
        Ground-truth accuracy under random-jitter motion.
        SSIM and RMSD compare the image-based, Type-2 NUFFT-based,
        and Type-1 NUFFT-based simulations with the analytical phantom
        reference at mild, moderate, and severe motion levels.
        Random jitter is presented separately from Figure~9 to
        accommodate its wider metric ranges.
        Points indicate outliers. Statistical comparisons are reported in Table S3.
    }
    \label{fig:si-random-jitter}
\end{figure}

\begin{sidewaysfigure}[p]
    \centering
    \includegraphics[width=0.95\linewidth]{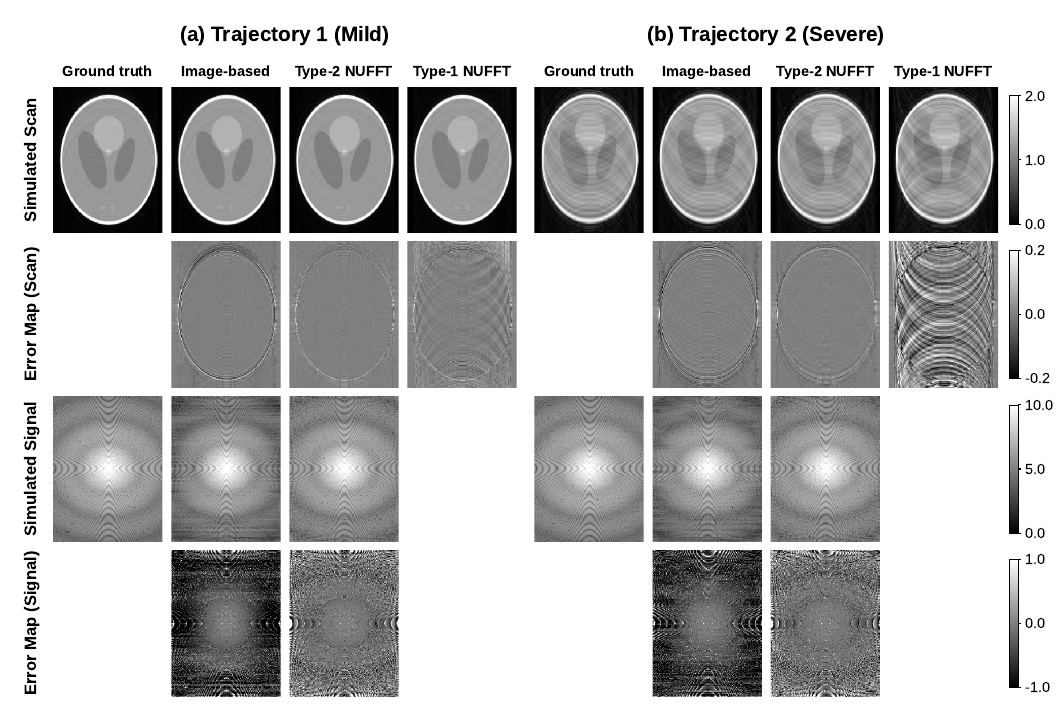}
    \caption{
        Image and signal comparisons for measured Trajectories~1
        (mild) and~2 (severe).
        Panels show reconstructed phantom images, image error maps,
        log-magnitude signals, and differences in log-magnitude signals.
        Differences are computed as the simulated result minus the
        analytical ground truth.
        Type-1 NUFFT signal comparisons are omitted because their
        non-uniform sampling locations prevent direct point-by-point
        comparison.
    }
    \label{fig:si-image-signal-errors}
\end{sidewaysfigure}

\begin{figure}[p]
    \centering
    \includegraphics[width=0.8\linewidth]{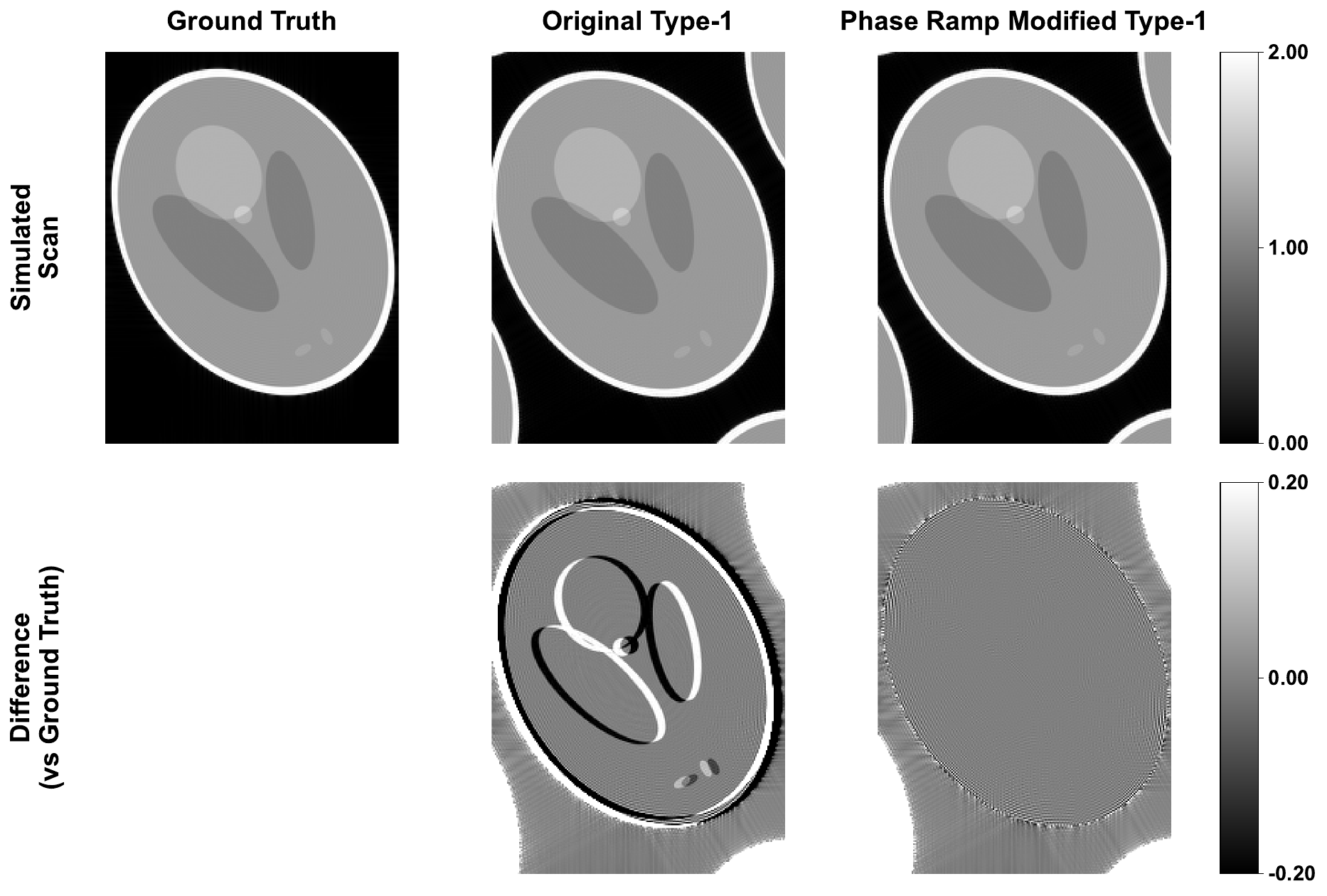}

    \caption{Demonstration that the original Type-1 NUFFT-based
    algorithm of Duffy et al. does not apply the intended translation
    correctly when rotational motion is also present. The simulations
    use constant motion parameters comprising a $30^\circ$ rotation
    about the $z$-axis and a $10$-mm translation along the $y$-axis.
    Owing to the non-conventional reconstruction, using the standard
    phase-ramp term
    $e^{-i2\pi\mathbf{k}_{\mathbf{m}}\cdot\mathbf{T}_t}$
    produces an incorrect translation. Replacing it with
    $e^{-i2\pi(\mathbf{R}_t\mathbf{k}_{\mathbf{m}})
    \cdot\mathbf{T}_t}$
    correctly applies the intended translation. The additional image
    copies at the corners result from using a Type-1 NUFFT for
    reconstruction.}
    \label{figS:type1_constant_motion}
\end{figure}

\begin{figure}[p]
    \centering

    \begin{subfigure}{\linewidth}
        \centering
        \includegraphics[width=0.7\linewidth]{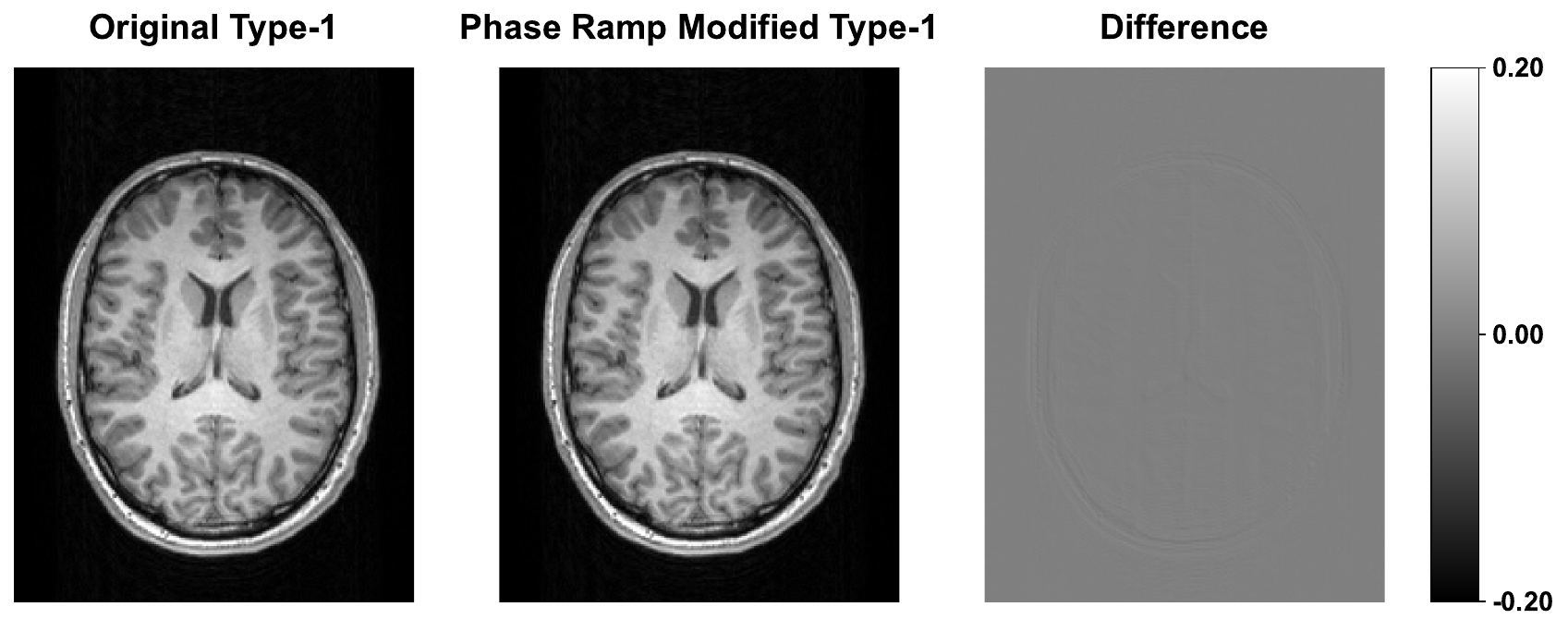}
        \caption{Trajectory 1}
    \end{subfigure}

    \begin{subfigure}{\linewidth}
        \centering
        \includegraphics[width=0.7\linewidth]{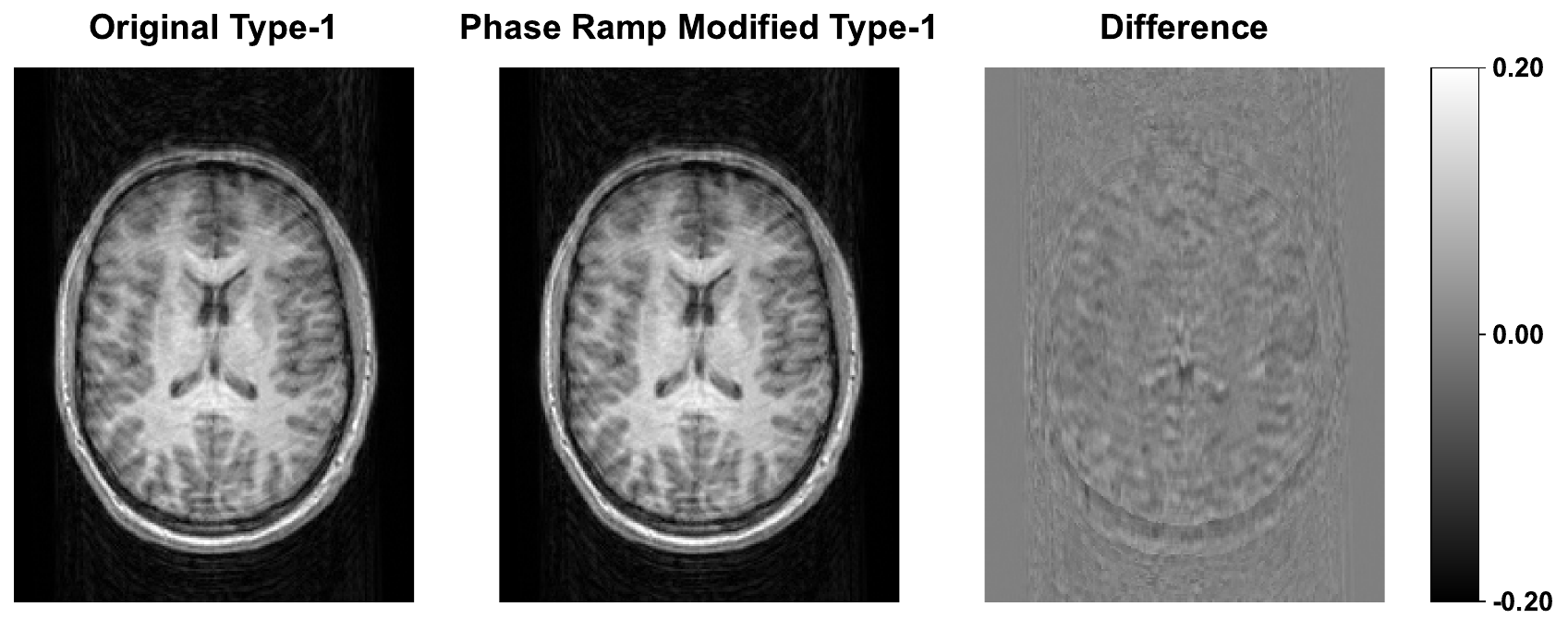}
        \caption{Trajectory 2}
    \end{subfigure}

    \caption{Comparison of the original and phase-ramp-modified
    Type-1 NUFFT-based algorithms on brain data. For realistic mild
    motion (Trajectory 1), the difference is negligible and is barely
    visible in the difference map. Under severe motion (Trajectory 2),
    the difference is more pronounced, although it remains difficult
    to identify by inspecting the simulated scans alone.}
    \label{figS:type1_brain_comparison}
\end{figure}

\begin{figure}[p]
    \centering

    \begin{subfigure}{\linewidth}
        \centering
        \includegraphics[width=0.6\linewidth]{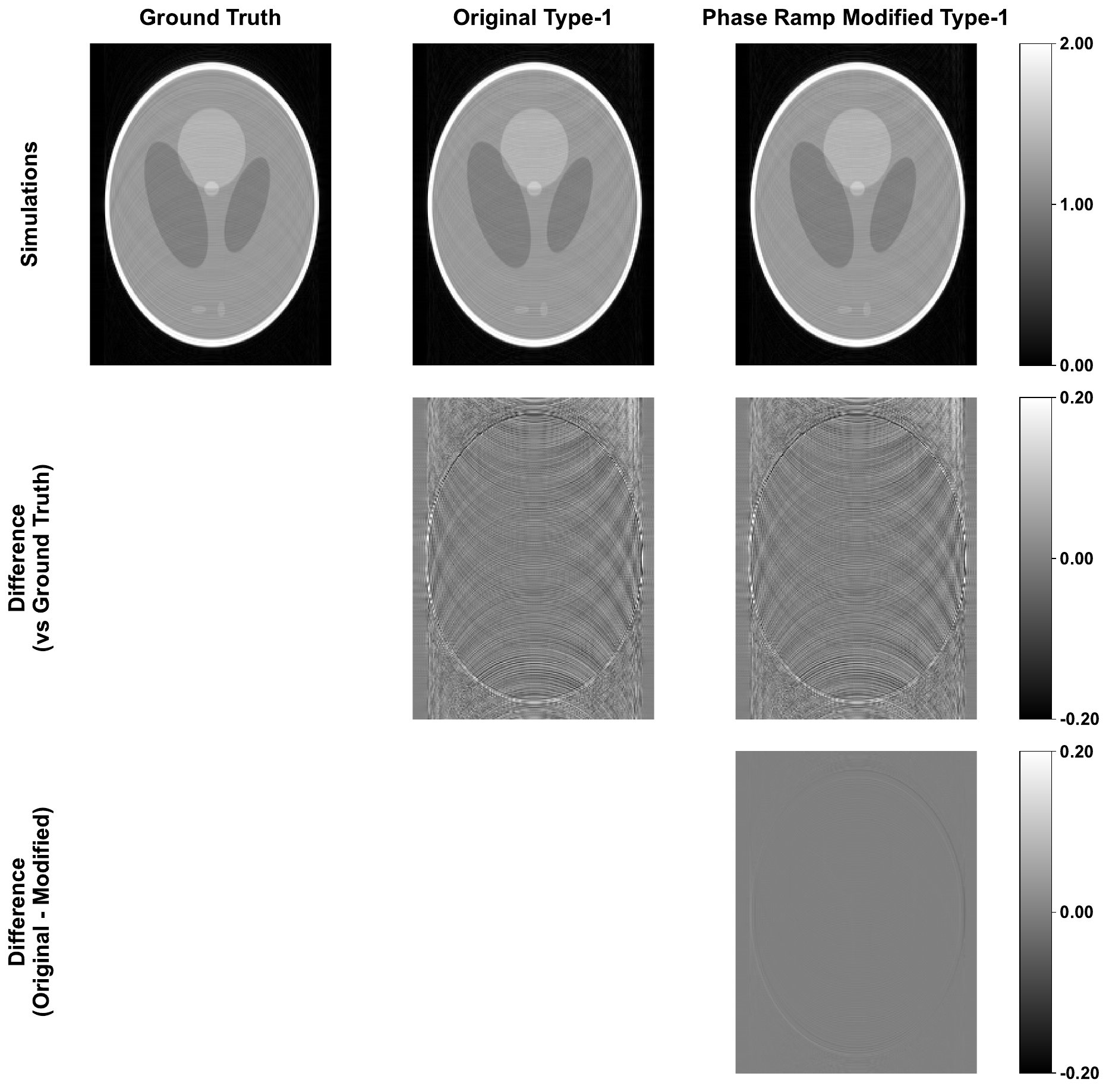}
        \caption{Trajectory 1}
    \end{subfigure}

    \begin{subfigure}{\linewidth}
        \centering
        \includegraphics[width=0.6\linewidth]{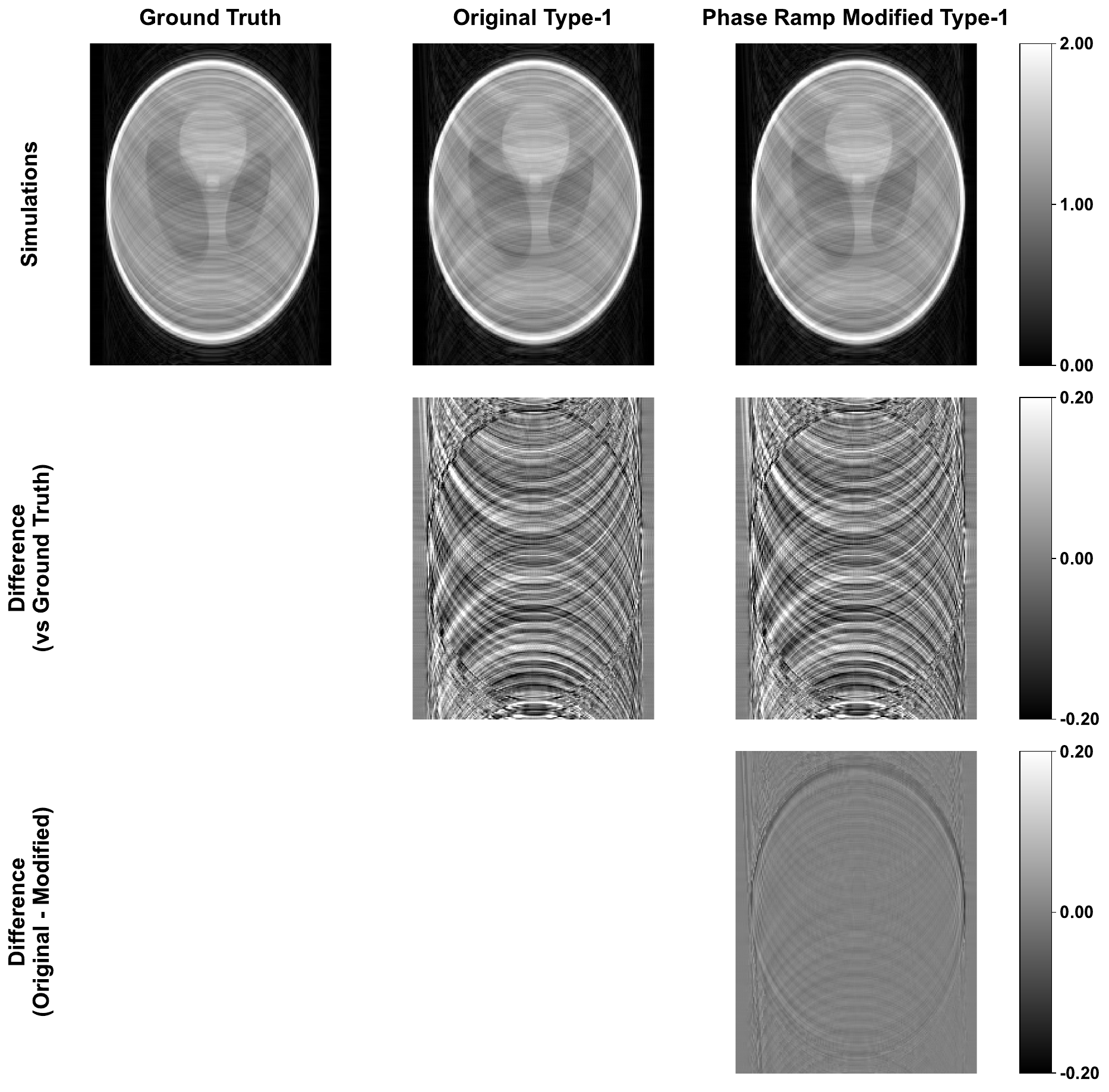}
        \caption{Trajectory 2}
    \end{subfigure}

    \caption{Comparison of the original and phase-ramp-modified
    Type-1 NUFFT-based algorithms on phantom data. The results show
    clear differences under severe motion (Trajectory 2). However,
    the deviation of the Type-1 NUFFT-based algorithm from the ground
    truth remains dominated by its fundamental theoretical
    inconsistency.}
    \label{figS:type1_phantom_comparison}
\end{figure}

\begin{table}[p]
\centering
\singlespacing
\small
\caption{Statistical comparisons of pairwise algorithm agreement for brain data, corresponding to Fig.~\ref{fig:si-brain-pairwise}. The three algorithm-pair groups were compared separately for SSIM and RMSD. Entries are Holm-adjusted \(p\)-values.}
\label{tab:si-brain-pairwise-statistics}
\setlength{\tabcolsep}{4pt}
\renewcommand{\arraystretch}{1.08}
\begin{tabular}{@{}llccc@{}}
\toprule
Motion type & Severity & \(A\) vs. \(B\) & \(A\) vs. \(C\) & \(B\) vs. \(C\) \\
\midrule
\multicolumn{5}{l}{\textit{SSIM}} \\
Slow drift & Mild & \(4.20\times10^{-8}\;{}^{\mathrm{t}}\) & \(3.28\times10^{-16}\;{}^{\mathrm{t}}\) & \(3.28\times10^{-16}\;{}^{\mathrm{t}}\) \\
Slow drift & Moderate & \(8.96\times10^{-8}\;{}^{\mathrm{t}}\) & \(5.01\times10^{-12}\;{}^{\mathrm{t}}\) & \(1.91\times10^{-6}\;{}^{\mathrm{W}}\) \\
Slow drift & Severe & \(5.85\times10^{-9}\;{}^{\mathrm{t}}\) & \(4.00\times10^{-5}\;{}^{\mathrm{t}}\) & \(5.50\times10^{-13}\;{}^{\mathrm{t}}\) \\
\addlinespace[2pt]
Spike motion & Mild & \(5.72\times10^{-6}\;{}^{\mathrm{W}}\) & \(5.72\times10^{-6}\;{}^{\mathrm{W}}\) & \(0.120\;{}^{\mathrm{t}}\)~\textbf{ns} \\
Spike motion & Moderate & \(1.62\times10^{-8}\;{}^{\mathrm{t}}\) & \(1.62\times10^{-8}\;{}^{\mathrm{t}}\) & \(1.28\times10^{-8}\;{}^{\mathrm{t}}\) \\
Spike motion & Severe & \(8.92\times10^{-10}\;{}^{\mathrm{t}}\) & \(7.41\times10^{-10}\;{}^{\mathrm{t}}\) & \(3.81\times10^{-6}\;{}^{\mathrm{W}}\) \\
\addlinespace[2pt]
Stepwise motion & Mild & \(3.81\times10^{-6}\;{}^{\mathrm{W}}\) & \(0.00486\;{}^{\mathrm{W}}\) & \(2.27\times10^{-19}\;{}^{\mathrm{t}}\) \\
Stepwise motion & Moderate & \(2.19\times10^{-9}\;{}^{\mathrm{t}}\) & \(2.26\times10^{-5}\;{}^{\mathrm{t}}\) & \(5.45\times10^{-15}\;{}^{\mathrm{t}}\) \\
Stepwise motion & Severe & \(2.20\times10^{-10}\;{}^{\mathrm{t}}\) & \(1.84\times10^{-7}\;{}^{\mathrm{t}}\) & \(5.24\times10^{-12}\;{}^{\mathrm{t}}\) \\
\addlinespace[2pt]
Random jitter & Mild & \(2.50\times10^{-21}\;{}^{\mathrm{t}}\) & \(8.44\times10^{-17}\;{}^{\mathrm{t}}\) & \(9.49\times10^{-28}\;{}^{\mathrm{t}}\) \\
Random jitter & Moderate & \(7.66\times10^{-23}\;{}^{\mathrm{t}}\) & \(1.01\times10^{-20}\;{}^{\mathrm{t}}\) & \(7.66\times10^{-23}\;{}^{\mathrm{t}}\) \\
Random jitter & Severe & \(5.04\times10^{-24}\;{}^{\mathrm{t}}\) & \(6.11\times10^{-23}\;{}^{\mathrm{t}}\) & \(2.06\times10^{-18}\;{}^{\mathrm{t}}\) \\
\midrule
\multicolumn{5}{l}{\textit{RMSD}} \\
Slow drift & Mild & \(1.53\times10^{-6}\;{}^{\mathrm{t}}\) & \(3.19\times10^{-18}\;{}^{\mathrm{t}}\) & \(3.19\times10^{-18}\;{}^{\mathrm{t}}\) \\
Slow drift & Moderate & \(8.61\times10^{-11}\;{}^{\mathrm{t}}\) & \(2.16\times10^{-14}\;{}^{\mathrm{t}}\) & \(5.53\times10^{-18}\;{}^{\mathrm{t}}\) \\
Slow drift & Severe & \(2.28\times10^{-11}\;{}^{\mathrm{t}}\) & \(4.85\times10^{-4}\;{}^{\mathrm{t}}\) & \(1.07\times10^{-8}\;{}^{\mathrm{t}}\) \\
\addlinespace[2pt]
Spike motion & Mild & \(5.72\times10^{-6}\;{}^{\mathrm{W}}\) & \(5.72\times10^{-6}\;{}^{\mathrm{W}}\) & \(5.72\times10^{-6}\;{}^{\mathrm{W}}\) \\
Spike motion & Moderate & \(3.81\times10^{-6}\;{}^{\mathrm{W}}\) & \(7.95\times10^{-10}\;{}^{\mathrm{t}}\) & \(3.81\times10^{-6}\;{}^{\mathrm{W}}\) \\
Spike motion & Severe & \(1.75\times10^{-12}\;{}^{\mathrm{t}}\) & \(1.44\times10^{-12}\;{}^{\mathrm{t}}\) & \(1.81\times10^{-13}\;{}^{\mathrm{t}}\) \\
\addlinespace[2pt]
Stepwise motion & Mild & \(1.91\times10^{-6}\;{}^{\mathrm{W}}\) & \(1.27\times10^{-19}\;{}^{\mathrm{t}}\) & \(8.56\times10^{-21}\;{}^{\mathrm{t}}\) \\
Stepwise motion & Moderate & \(1.91\times10^{-6}\;{}^{\mathrm{W}}\) & \(5.77\times10^{-7}\;{}^{\mathrm{t}}\) & \(3.17\times10^{-15}\;{}^{\mathrm{t}}\) \\
Stepwise motion & Severe & \(2.51\times10^{-9}\;{}^{\mathrm{t}}\) & \(0.0241\;{}^{\mathrm{t}}\) & \(2.86\times10^{-10}\;{}^{\mathrm{t}}\) \\
\addlinespace[2pt]
Random jitter & Mild & \(1.82\times10^{-20}\;{}^{\mathrm{t}}\) & \(7.42\times10^{-7}\;{}^{\mathrm{t}}\) & \(1.53\times10^{-24}\;{}^{\mathrm{t}}\) \\
Random jitter & Moderate & \(2.86\times10^{-21}\;{}^{\mathrm{t}}\) & \(1.20\times10^{-17}\;{}^{\mathrm{t}}\) & \(2.14\times10^{-25}\;{}^{\mathrm{t}}\) \\
Random jitter & Severe & \(4.73\times10^{-22}\;{}^{\mathrm{t}}\) & \(1.08\times10^{-20}\;{}^{\mathrm{t}}\) & \(7.49\times10^{-18}\;{}^{\mathrm{t}}\) \\
\bottomrule
\end{tabular}
\par\vspace{5pt}
\begin{minipage}{\textwidth}
\footnotesize
\(A\): Type-2 NUFFT versus image-based; \(B\): Type-1 NUFFT versus image-based; \(C\): Type-1 NUFFT versus Type-2 NUFFT. Comparisons are between these pairwise-agreement groups, not against analytical ground truth.\par\smallskip
\(^{\mathrm{t}}\) Two-sided paired \(t\)-test;
\(^{\mathrm{W}}\) two-sided Wilcoxon signed-rank test.
Each comparison included 20 observations matched by trajectory.
A paired \(t\)-test was used when the Shapiro--Wilk test of the paired
differences did not reject normality (\(p\geq0.05\)); otherwise,
a Wilcoxon signed-rank test was used. Holm correction was applied across
the three comparisons within each motion type, severity level, and metric.
Bold \textbf{ns} denotes \(p_{\mathrm{Holm}}\geq0.05\).
Values are reported to three significant figures; significance was
determined using the unrounded values.
\end{minipage}
\end{table}

\begin{table}[p]
\centering
\singlespacing
\small
\caption{Statistical comparisons of pairwise algorithm agreement for the digital phantom, corresponding to Fig.~\ref{fig:si-phantom-pairwise}. The three algorithm-pair groups were compared separately for SSIM and RMSD. Entries are Holm-adjusted \(p\)-values.}
\label{tab:si-phantom-pairwise-statistics}
\setlength{\tabcolsep}{4pt}
\renewcommand{\arraystretch}{1.08}
\begin{tabular}{@{}llccc@{}}
\toprule
Motion type & Severity & \(A\) vs. \(B\) & \(A\) vs. \(C\) & \(B\) vs. \(C\) \\
\midrule
\multicolumn{5}{l}{\textit{SSIM}} \\
Slow drift & Mild & \(0.00295\;{}^{\mathrm{t}}\) & \(4.37\times10^{-14}\;{}^{\mathrm{t}}\) & \(7.87\times10^{-14}\;{}^{\mathrm{t}}\) \\
Slow drift & Moderate & \(1.86\times10^{-8}\;{}^{\mathrm{t}}\) & \(5.61\times10^{-9}\;{}^{\mathrm{t}}\) & \(1.21\times10^{-18}\;{}^{\mathrm{t}}\) \\
Slow drift & Severe & \(9.38\times10^{-11}\;{}^{\mathrm{t}}\) & \(1.07\times10^{-7}\;{}^{\mathrm{t}}\) & \(3.31\times10^{-14}\;{}^{\mathrm{t}}\) \\
\addlinespace[2pt]
Spike motion & Mild & \(9.13\times10^{-8}\;{}^{\mathrm{t}}\) & \(7.93\times10^{-8}\;{}^{\mathrm{t}}\) & \(2.72\times10^{-8}\;{}^{\mathrm{t}}\) \\
Spike motion & Moderate & \(1.68\times10^{-10}\;{}^{\mathrm{t}}\) & \(1.27\times10^{-10}\;{}^{\mathrm{t}}\) & \(1.80\times10^{-15}\;{}^{\mathrm{t}}\) \\
Spike motion & Severe & \(6.20\times10^{-11}\;{}^{\mathrm{t}}\) & \(6.00\times10^{-11}\;{}^{\mathrm{t}}\) & \(1.92\times10^{-19}\;{}^{\mathrm{t}}\) \\
\addlinespace[2pt]
Stepwise motion & Mild & \(3.81\times10^{-6}\;{}^{\mathrm{W}}\) & \(0.651\;{}^{\mathrm{t}}\)~\textbf{ns} & \(1.40\times10^{-17}\;{}^{\mathrm{t}}\) \\
Stepwise motion & Moderate & \(1.51\times10^{-9}\;{}^{\mathrm{t}}\) & \(1.38\times10^{-7}\;{}^{\mathrm{t}}\) & \(1.33\times10^{-11}\;{}^{\mathrm{t}}\) \\
Stepwise motion & Severe & \(6.96\times10^{-10}\;{}^{\mathrm{t}}\) & \(8.37\times10^{-9}\;{}^{\mathrm{t}}\) & \(4.23\times10^{-7}\;{}^{\mathrm{t}}\) \\
\addlinespace[2pt]
Random jitter & Mild & \(8.53\times10^{-21}\;{}^{\mathrm{t}}\) & \(1.66\times10^{-17}\;{}^{\mathrm{t}}\) & \(6.74\times10^{-24}\;{}^{\mathrm{t}}\) \\
Random jitter & Moderate & \(1.19\times10^{-21}\;{}^{\mathrm{t}}\) & \(2.27\times10^{-20}\;{}^{\mathrm{t}}\) & \(1.19\times10^{-21}\;{}^{\mathrm{t}}\) \\
Random jitter & Severe & \(1.55\times10^{-22}\;{}^{\mathrm{t}}\) & \(4.82\times10^{-22}\;{}^{\mathrm{t}}\) & \(1.99\times10^{-17}\;{}^{\mathrm{t}}\) \\
\midrule
\multicolumn{5}{l}{\textit{RMSD}} \\
Slow drift & Mild & \(3.17\times10^{-5}\;{}^{\mathrm{t}}\) & \(1.78\times10^{-18}\;{}^{\mathrm{t}}\) & \(1.78\times10^{-18}\;{}^{\mathrm{t}}\) \\
Slow drift & Moderate & \(5.13\times10^{-11}\;{}^{\mathrm{t}}\) & \(1.08\times10^{-15}\;{}^{\mathrm{t}}\) & \(7.18\times10^{-19}\;{}^{\mathrm{t}}\) \\
Slow drift & Severe & \(1.22\times10^{-11}\;{}^{\mathrm{t}}\) & \(0.0173\;{}^{\mathrm{t}}\) & \(7.94\times10^{-11}\;{}^{\mathrm{t}}\) \\
\addlinespace[2pt]
Spike motion & Mild & \(3.81\times10^{-6}\;{}^{\mathrm{W}}\) & \(3.81\times10^{-6}\;{}^{\mathrm{W}}\) & \(2.15\times10^{-12}\;{}^{\mathrm{t}}\) \\
Spike motion & Moderate & \(2.97\times10^{-9}\;{}^{\mathrm{t}}\) & \(2.77\times10^{-9}\;{}^{\mathrm{t}}\) & \(1.14\times10^{-13}\;{}^{\mathrm{t}}\) \\
Spike motion & Severe & \(7.42\times10^{-10}\;{}^{\mathrm{t}}\) & \(5.97\times10^{-10}\;{}^{\mathrm{t}}\) & \(1.91\times10^{-6}\;{}^{\mathrm{W}}\) \\
\addlinespace[2pt]
Stepwise motion & Mild & \(1.91\times10^{-6}\;{}^{\mathrm{W}}\) & \(2.59\times10^{-18}\;{}^{\mathrm{t}}\) & \(4.32\times10^{-20}\;{}^{\mathrm{t}}\) \\
Stepwise motion & Moderate & \(2.10\times10^{-8}\;{}^{\mathrm{t}}\) & \(5.10\times10^{-4}\;{}^{\mathrm{t}}\) & \(5.36\times10^{-14}\;{}^{\mathrm{t}}\) \\
Stepwise motion & Severe & \(3.18\times10^{-9}\;{}^{\mathrm{t}}\) & \(0.0125\;{}^{\mathrm{t}}\) & \(2.08\times10^{-10}\;{}^{\mathrm{t}}\) \\
\addlinespace[2pt]
Random jitter & Mild & \(4.10\times10^{-20}\;{}^{\mathrm{t}}\) & \(2.95\times10^{-11}\;{}^{\mathrm{t}}\) & \(4.04\times10^{-25}\;{}^{\mathrm{t}}\) \\
Random jitter & Moderate & \(8.24\times10^{-21}\;{}^{\mathrm{t}}\) & \(2.49\times10^{-18}\;{}^{\mathrm{t}}\) & \(6.89\times10^{-26}\;{}^{\mathrm{t}}\) \\
Random jitter & Severe & \(2.36\times10^{-21}\;{}^{\mathrm{t}}\) & \(1.62\times10^{-20}\;{}^{\mathrm{t}}\) & \(5.78\times10^{-17}\;{}^{\mathrm{t}}\) \\
\bottomrule
\end{tabular}
\par\vspace{5pt}
\begin{minipage}{\textwidth}
\footnotesize
\(A\): Type-2 NUFFT versus image-based; \(B\): Type-1 NUFFT versus image-based; \(C\): Type-1 NUFFT versus Type-2 NUFFT. Comparisons are between these pairwise-agreement groups, not against analytical ground truth.\par\smallskip
\(^{\mathrm{t}}\) Two-sided paired \(t\)-test;
\(^{\mathrm{W}}\) two-sided Wilcoxon signed-rank test.
Each comparison included 20 observations matched by trajectory.
A paired \(t\)-test was used when the Shapiro--Wilk test of the paired
differences did not reject normality (\(p\geq0.05\)); otherwise,
a Wilcoxon signed-rank test was used. Holm correction was applied across
the three comparisons within each motion type, severity level, and metric.
Bold \textbf{ns} denotes \(p_{\mathrm{Holm}}\geq0.05\).
Values are reported to three significant figures; significance was
determined using the unrounded values.
\end{minipage}
\end{table}

\begin{table}[p]
\centering
\singlespacing
\small
\caption{Statistical comparisons of simulation accuracy relative to analytical ground truth, corresponding to Figure 9 in the main manuscript and Figure S8 in the Supporting Information. SSIM and RMSD against ground truth were compared between simulation methods. Entries are Holm-adjusted \(p\)-values.}
\label{tab:si-ground-truth-statistics}
\setlength{\tabcolsep}{4pt}
\renewcommand{\arraystretch}{1.08}
\begin{tabular}{@{}llccc@{}}
\toprule
Motion type & Severity & Type-1 vs. Type-2 & Type-1 vs. image & Type-2 vs. image \\
\midrule
\multicolumn{5}{l}{\textit{SSIM}} \\
Slow drift & Mild & \(0.0105\;{}^{\mathrm{t}}\) & \(1.30\times10^{-5}\;{}^{\mathrm{t}}\) & \(1.14\times10^{-4}\;{}^{\mathrm{t}}\) \\
Slow drift & Moderate & \(2.76\times10^{-9}\;{}^{\mathrm{t}}\) & \(6.34\times10^{-5}\;{}^{\mathrm{t}}\) & \(2.36\times10^{-4}\;{}^{\mathrm{t}}\) \\
Slow drift & Severe & \(9.41\times10^{-12}\;{}^{\mathrm{t}}\) & \(2.79\times10^{-10}\;{}^{\mathrm{t}}\) & \(1.92\times10^{-7}\;{}^{\mathrm{t}}\) \\
\addlinespace[2pt]
Spike motion & Mild & \(1.70\times10^{-8}\;{}^{\mathrm{t}}\) & \(3.45\times10^{-8}\;{}^{\mathrm{t}}\) & \(6.55\times10^{-11}\;{}^{\mathrm{t}}\) \\
Spike motion & Moderate & \(5.09\times10^{-11}\;{}^{\mathrm{t}}\) & \(5.09\times10^{-11}\;{}^{\mathrm{t}}\) & \(8.28\times10^{-17}\;{}^{\mathrm{t}}\) \\
Spike motion & Severe & \(3.33\times10^{-11}\;{}^{\mathrm{t}}\) & \(3.33\times10^{-11}\;{}^{\mathrm{t}}\) & \(3.74\times10^{-17}\;{}^{\mathrm{t}}\) \\
\addlinespace[2pt]
Stepwise motion & Mild & \(5.72\times10^{-6}\;{}^{\mathrm{W}}\) & \(5.72\times10^{-6}\;{}^{\mathrm{W}}\) & \(0.00254\;{}^{\mathrm{t}}\) \\
Stepwise motion & Moderate & \(1.97\times10^{-10}\;{}^{\mathrm{t}}\) & \(7.90\times10^{-9}\;{}^{\mathrm{t}}\) & \(1.91\times10^{-6}\;{}^{\mathrm{W}}\) \\
Stepwise motion & Severe & \(2.07\times10^{-11}\;{}^{\mathrm{t}}\) & \(6.00\times10^{-10}\;{}^{\mathrm{t}}\) & \(1.91\times10^{-6}\;{}^{\mathrm{W}}\) \\
\addlinespace[2pt]
Random jitter & Mild & \(1.44\times10^{-20}\;{}^{\mathrm{t}}\) & \(2.89\times10^{-17}\;{}^{\mathrm{t}}\) & \(5.30\times10^{-19}\;{}^{\mathrm{t}}\) \\
Random jitter & Moderate & \(1.58\times10^{-22}\;{}^{\mathrm{t}}\) & \(4.90\times10^{-20}\;{}^{\mathrm{t}}\) & \(1.46\times10^{-23}\;{}^{\mathrm{t}}\) \\
Random jitter & Severe & \(1.46\times10^{-23}\;{}^{\mathrm{t}}\) & \(4.05\times10^{-22}\;{}^{\mathrm{t}}\) & \(3.29\times10^{-27}\;{}^{\mathrm{t}}\) \\
\midrule
\multicolumn{5}{l}{\textit{RMSD}} \\
Slow drift & Mild & \(1.91\times10^{-6}\;{}^{\mathrm{W}}\) & \(7.89\times10^{-17}\;{}^{\mathrm{t}}\) & \(7.89\times10^{-17}\;{}^{\mathrm{t}}\) \\
Slow drift & Moderate & \(3.81\times10^{-6}\;{}^{\mathrm{W}}\) & \(1.16\times10^{-19}\;{}^{\mathrm{t}}\) & \(3.81\times10^{-6}\;{}^{\mathrm{W}}\) \\
Slow drift & Severe & \(1.90\times10^{-13}\;{}^{\mathrm{t}}\) & \(0.516\;{}^{\mathrm{t}}\)~\textbf{ns} & \(3.60\times10^{-29}\;{}^{\mathrm{t}}\) \\
\addlinespace[2pt]
Spike motion & Mild & \(2.78\times10^{-9}\;{}^{\mathrm{t}}\) & \(5.45\times10^{-9}\;{}^{\mathrm{t}}\) & \(1.99\times10^{-14}\;{}^{\mathrm{t}}\) \\
Spike motion & Moderate & \(3.04\times10^{-9}\;{}^{\mathrm{t}}\) & \(3.04\times10^{-9}\;{}^{\mathrm{t}}\) & \(1.43\times10^{-17}\;{}^{\mathrm{t}}\) \\
Spike motion & Severe & \(8.01\times10^{-10}\;{}^{\mathrm{t}}\) & \(8.01\times10^{-10}\;{}^{\mathrm{t}}\) & \(5.64\times10^{-21}\;{}^{\mathrm{t}}\) \\
\addlinespace[2pt]
Stepwise motion & Mild & \(1.91\times10^{-6}\;{}^{\mathrm{W}}\) & \(9.00\times10^{-20}\;{}^{\mathrm{t}}\) & \(3.48\times10^{-22}\;{}^{\mathrm{t}}\) \\
Stepwise motion & Moderate & \(2.96\times10^{-10}\;{}^{\mathrm{t}}\) & \(1.01\times10^{-5}\;{}^{\mathrm{t}}\) & \(1.52\times10^{-23}\;{}^{\mathrm{t}}\) \\
Stepwise motion & Severe & \(1.50\times10^{-10}\;{}^{\mathrm{t}}\) & \(0.163\;{}^{\mathrm{t}}\)~\textbf{ns} & \(4.52\times10^{-20}\;{}^{\mathrm{t}}\) \\
\addlinespace[2pt]
Random jitter & Mild & \(4.26\times10^{-21}\;{}^{\mathrm{t}}\) & \(2.48\times10^{-8}\;{}^{\mathrm{t}}\) & \(7.49\times10^{-28}\;{}^{\mathrm{t}}\) \\
Random jitter & Moderate & \(3.03\times10^{-22}\;{}^{\mathrm{t}}\) & \(1.32\times10^{-16}\;{}^{\mathrm{t}}\) & \(3.95\times10^{-35}\;{}^{\mathrm{t}}\) \\
Random jitter & Severe & \(4.45\times10^{-23}\;{}^{\mathrm{t}}\) & \(1.13\times10^{-19}\;{}^{\mathrm{t}}\) & \(6.22\times10^{-35}\;{}^{\mathrm{t}}\) \\
\bottomrule
\end{tabular}
\par\vspace{5pt}
\begin{minipage}{\textwidth}
\footnotesize
Type-1 and Type-2 denote NUFFT-based simulation methods; image denotes the image-based method. These tests compare ground-truth agreement, rather than direct pairwise agreement between simulated images.\par\smallskip
\(^{\mathrm{t}}\) Two-sided paired \(t\)-test;
\(^{\mathrm{W}}\) two-sided Wilcoxon signed-rank test.
Each comparison included 20 observations matched by trajectory.
A paired \(t\)-test was used when the Shapiro--Wilk test of the paired
differences did not reject normality (\(p\geq0.05\)); otherwise,
a Wilcoxon signed-rank test was used. Holm correction was applied across
the three comparisons within each motion type, severity level, and metric.
Bold \textbf{ns} denotes \(p_{\mathrm{Holm}}\geq0.05\).
Values are reported to three significant figures; significance was
determined using the unrounded values.
\end{minipage}
\end{table}